\documentstyle[12pt,aasms4,natbib]{article}
\newcommand{\I}{$I$~}
\newcommand{\etal}{$et \; al.$~}

\def\lesssim{\mathrel{\hbox{\rlap{\hbox{%
 \lower4pt\hbox{$\sim$}}}\hbox{$<$}}}}
\def\gtrsim{\mathrel{\hbox{\rlap{\hbox{%
 \lower4pt\hbox{$\sim$}}}\hbox{$>$}}}}

\def\arcmin{\hbox{$^\prime \,$}}
\def\arcsec{\hbox{$^{\prime\prime \, }$}}

\def\farcs{\hbox{$.\!\!^{\prime\prime}$}}

\begin{document}

\title{High Redshift Supernovae in the Hubble Deep Field\altaffilmark{1}}
\author{Ronald L. Gilliland\altaffilmark{2}, Peter E. Nugent\altaffilmark{3},
M. M. Phillips\altaffilmark{4}}

\lefthead{Gilliland et al.}
\righthead{High-z SNe in the HDF}

\altaffiltext{1}{Based on observations with the NASA/ESA {\em Hubble 
Space Telescope}, obtained at the Space Telescope Science Institute,
which is operated by AURA, Inc., under NASA contract NAS 5-26555.}
\altaffiltext{2}{Space Telescope Science Institute, 3700 San Martin
Drive, Baltimore, MD 21218; gillil@stsci.edu}
\altaffiltext{3}{E.O. Lawrence Berkeley National Laboratory, 1 Cyclotron
Road, MS 50-232, Berkeley, CA  94720; penugent@lbl.gov}
\altaffiltext{4}{Carnegie Institution of Washington, Las Campanas Observatory,
Casilla 601, La Serena, Chile; mmp@ociw.edu}

\begin{abstract}
Two supernovae detected in the Hubble Deep Field (HDF) using the
original December 1995 epoch and data from a shorter (63000 s in
F814W) December 1997 visit with {\em HST} are discussed.  The
supernovae (SNe) are both associated with distinct galaxies at
redshifts of 0.95 (spectroscopic) from \cite{cohen_96} and 1.32
(photometric) from the work of \cite{soto98}.  These redshifts are
near, in the case of 0.95, and well beyond for 1.32 the greatest
distance reported previously for SNe.  We show that our observations
are sensitive to supernovae to z $\lesssim$ 1.8 in either
epoch for an event near peak brightness.  Detailed simulations are
discussed that quantify the level at which false events from our
search phase would start to arise, and the completeness of our search
as a function of both SN brightness and host galaxy redshift.  The
number of Type Ia and Type II SNe expected as a function of redshift in the two HDF epochs
are discussed in relation to several published predictions and our own
detailed calculations.  A mean detection frequency of one SN per epoch
for the small HDF area is consistent with expectations from current
theory.
\end{abstract}


\section{Introduction}

Supernovae provide sensitive probes for conditions in the early
universe.  Type II supernovae (SNe~II) events arising from final
gravitational collapse and rebound in massive stars reflect the star
formation rate (SFR) with essentially no lag time for the high end of
the initial mass function (IMF) and their detection may allow
quantification of the SFR even in galaxies which are too faint for
direct imaging.  Type Ia supernovae (SNe~Ia) are the brightest
supernovae (with luminous Type Ic events also at the top end of brightness distributions) and are established as standard candles on both theoretical
and observational grounds.
The extent to which Type Ia SNe deviate from being perfect standard candles in 
terms of intrinsic magnitude scatter following corrections from decline
rate and spectral intensity distribution correlations, as well as possible systematics
with look back time remain topics of vigorous investigation.
SNe~Ia are thought to arise from
thermonuclear runaways in White Dwarfs (WDs) passing the Chandrasekhar
mass limit as a result of either double degenerate mergers or mass
transfer and accretion in a close binary system.  The SN~Ia rates are
therefore less directly connected to the SFR than for SNe~II, but
nonetheless with a lag necessary to first form WDs via stellar
evolution, then time for the WD to grow in mass, their rates will
reflect the rate of star formation in the early universe.

Supernovae are also responsible to first order for the overall
chemical evolution of the Universe. SNe~II are the primary source of
metals below the Iron peak and provide early heavy element creation
from the most massive stars.  SNe~Ia are thought to provide the
primary source of Iron peak elements.  The relative frequency of these
SNe types (SNe~II coming first followed by a later relative increase
in the SN~Ia rates) can therefore explain in a simple way the overall
increase of metals with time in an evolving universe, and also account
for the relative overabundance of Oxygen to Iron seen in the early
universe and old stellar populations.

The detection and study of SNe are of obvious importance both for
their utility as cosmological probes, and also for their fundamental
role in determining chemical evolution.  Using specially crafted,
two-epoch, ground-based searches (e.g., \cite{42SNe_98};
\cite{Schmidt_98}) many SN~Ia events are now routinely detected
near z $\sim$ 0.5, and a few to z $\sim$ 1.0 (\cite{98hst_hiz}) and even beyond have been
found \cite[]{albinoni_iau}.  Using SNe~Ia as standard candles yields
a cosmology with low matter density and a non-zero cosmological
constant \cite[]{42SNe_98,riess_scoop98}.

SNe~Ia near maximum light have a flux maximum near 400-500 nm and a
steep flux decline below about 260 nm \cite[]{kir92a}.  Programs
designed for efficient detection of distant SNe are constrained to
redshifts of 1+z $\sim$ filter center wavelength/260 nm.  For the
longest wavelength filter, F814W, used for the HDF this corresponds to
z $\sim$ 2.0, the effective limit of our two-epoch SN search.  At z
$\sim$ 1.7 a normal SN~Ia will have a peak \I - band (we will use
Johnson-Cousins magnitudes throughout) magnitude of $m_I \sim$ 26.5
which is already beyond the detection limit based on point spread
function shape alone (\cite{flynn96}) that point source versus diffuse
object discrimination can be made for even the uniquely deep HDF.
Detection and detailed study of SNe~Ia through their first occurrence
(perhaps z $\sim$ 3-4 given a reasonable delay time from the Big Bang)
in the early universe is one of the key drivers for the Next
Generation Space Telescope (NGST) planning (\cite{madau_98}).  Our
WFPC2-{\em HST} search uses the existing F814W deep image of the HDF
and an additional (less) deep F814W HDF image two years later
providing two independent epochs in which SNe can be searched for.
The project goal is to quantify the utility of {\em HST} for
conducting searches for cosmological SNe and to return information on
the SN rate to z $\sim$ 2 in the early universe.  Since this initial
project does not allow for obtaining verification spectra to classify
individual events by type, nor is a substantial time base during
either epoch of observation available to define a light curve,
quantitative application to testing cosmology parameters is not
expected.

The observations of the original HDF and our second epoch imaging will
be briefly reviewed in \S 2.  Since our data reductions differ from
those used for the standard HDF releases \cite[]{williams_96},
discussion is provided in \S 3 to document these basics.  The SN
detected in the HDF will be discussed in detail in \S 4 along with
quantification of completeness limits and false signal occurrence.
Detectability of SNe with redshift, given our data characteristics and
intrinsic properties of SNe~Ia and SNe~II is developed in \S 5 along
with a comparison with several recent theoretical predictions.
Discussion of cosmological and more mundane implications is given in
the \S 6 summary.
\section{Observations}

With goals of pushing {\em HST} capabilities (pre-NICMOS for this
Cycle 6 program) for detection of high-z SN, and detecting a few
events with concomitant implications for SN rates at high z we
obtained second epoch observations of the HDF in the F814W filter.
Although the F606W HDF images are intrinsically deeper, F814W provides
a greater sensitivity for high-z SN events nearly in proportion to the
relative filter wavelengths as a result of the steep drop off in UV
flux (this has been shown with detailed K-correction evaluations).
SNe will be detected as point-like
sources that are visible in either, but not both epochs.

To maximize mutual areal coverage, and to minimize problems with image
differences across epochs for these under-sampled data, the second
epoch data were obtained at an {\em HST} orient as closely matching
the original as possible.

In order to obtain critical sampling for the WF pixel size of 0\farcs1
with inherently undersampled data the original epoch F814W exposures
were dithered to 10 sub-pixel pointings (9 by design, one extra as a
result of a guide star acquisition problem in one visit).  The
sub-pixel offsets were linked with large scale offsets over $\sim \pm$
2\arcsec to help in averaging out spatial imperfections in the CCD
response.  At each dither position 4 to 6 exposures were taken and
then cosmic rays were eliminated as multi-sigma deviations across
these short stacks (in the standard reductions described in
\cite{williams_96}) at common pointings.

A constraint of this SN program is to not only detect objects to well
quantified completeness limits, but to do so with an assurance that
{\bf no} false events from either noise fluctuations (Poisson, cosmic
rays) or CCD defects such as hot pixels are included.  Robust image
subtractions require excellent sampling implying a need for at least
as many sub-pixel dithers in the second epoch observations as in the
original.  Robust control against hot pixels (in undersampled data a
hot pixel is difficult to distinguish from a faint point source, and
over the course of these observations many new hot pixels are created)
argues for a large number of dithers to independent pixels.  For these
second epoch observations (18 orbits in the continuous viewing zone)
one independent dither offset per orbit was used.  In order to keep
readout noise well below sky only two F814W exposures (1600s, 1900s)
were taken per orbit, thus only two (or in the case of F300W exposures
taken during the bright phase of each orbit, one) exposures at any one
pointing.  Although for many projects cosmic ray elimination using
pairs of exposures would be acceptable, it would be a disaster for
this project since many spurious signals would arise from chance
coincidence of cosmic rays across image pairs.  We therefore required
code capable of eliminating cosmic rays across the full image stacks
simultaneously, even if all of the exposures had unique pointings.

Table~\ref{tab1} summarizes the original epoch F814W data and the
second epoch F814W and F300W data.
The F300W images are used as darks to define hot pixels for the purposes
of these observations, and extend the archival data base.  The F300W
images do not have a magnitude limit useful for constraints on high-z SNe.
The second epoch for F814W was
731.9 days after the first, or 2.00 years.  For both epochs the F814W
data was acquired in generally dark, and nearly equal conditions (only
$\sim$20\% of frames show any evidence of scattered light).  During
the original epoch a subset of 4 frames had a 4.5\arcmin rotation with
respect to the others; all of the second epoch data was rotated by
8.8\arcmin with respect to the bulk of the original.  Total exposure
time in the second epoch F814W data is 51\% of that obtained in the
original HDF.

\placetable{tab1}

\section{Data Reductions}

For the data reductions described in this section the following goals
apply: (1) Derive an over-sampled, summed image for each epoch of
F814W data.  (2) Use optimal weights to minimize background noise.
(3) Use precise knowledge of relative frame-to-frame offsets and an
algorithm that provides sharp, clean point spread functions.  (4)
Eliminate both cosmic rays and hot pixels such that any residual
signals from these are well below claimed SN detection levels.  (5)
Correct for the WFPC2 scattered light pattern in any affected frames.
(6) Construct the second epoch image on exactly the same sky-reference
grid as defined by external galaxies as was used for the first epoch
solution.  (7) Provide an error array for each epoch that accurately
reflects relative errors.  Successfully reaching these goals will
enable SN searches by simply taking differences of the two epoch
images and searching for point-like sources at several sigma above the
expected background fluctuations.

Several of the data reduction steps discussed next are not
independent, for example derivation of an image model and good cosmic
ray rejection requires knowledge of the image registration, and an
accurate derivation of the image registrations requires prior
knowledge of the ``true'' image and good cosmic ray elimination.
Fortunately the process is convergent, adopting a decent guess for the
image registrations (from the commanded dithers, or from ``off-line''
measurement of a few primary sources) provides sufficient accuracy for
a first round of image model creation and cosmic ray elimination.  The
initial image model then supports improved derivation of relative
registrations.

The reduction steps discussed below all start with the pipeline
calibrated images (for the original epoch the recalibrations as
discussed in \cite{williams_96}).  We start by rotating data for each
CCD by multiples of 90 degrees to align on the sky with WF4.  In this
initial step the data are also scaled to e- by multiplying by the gain
($\sim$7 e-/DN) for each CCD, the resulting images are stored as i*2
data to save storage (no degradation of signal to noise results from
this step).  A global sky level is evaluated for each frame as the
first moment position of a histogram within 3 $\sigma$ of the peak
generated from the central 600$\times$600 pixels for each frame, this
global value is then subtracted from each image.
\subsection{Image Model and Cosmic Ray Rejection}

Elimination of cosmic rays from Wide Field Camera data containing
stellar images and for which even minor frame-to-frame offsets exist
is problematic (most of the HDF area is rather easier to deal with
since most structures are slowly varying at the pixel scale).  The
sharp PSFs lead to variations at a given pixel of recorded flux near
stellar cores that vary by a full factor of 10 with relative dithers
of one pixel.  In order to retain full sensitivity to detecting cosmic
rays at a few sigma above a pixel's expected noise level requires
knowledge of how the intrinsic intensity changes for a given pixel
correlate with the relative change of pixel to sky mapping
frame-to-frame: this is the definition of image model (see discussion
in \cite{gill95}).

Given a series of independent images in time with sub-pixel offsets
frame-to-frame of $\delta x_{i,j}(t) , \: \delta y_{i,j}(t)$ the image
model is:

\begin{equation}
I_{i,j}(t)\,=\, f[ \delta x_{i,j}(t) , \: \delta y_{i,j}(t)]
\end{equation}

\noindent
where the function $f$ is chosen to have enough variation to match the
rapid change of $I_{i,j}$ experienced in under-sampled data.  We have
used up to a third-order polynomial (10 terms) in both $x$ and $y$ to
yield a good model intensity.  In this approach the expected CCD image
of the sky at any arbitrary sub-pixel offset, as sampled by the broad
CCD pixels, can be calculated as a function of the relative sub-pixel
(integer shifts are factored out of this representation) $x,\,y$
offset.  It is then a simple matter to flag and eliminate cosmic rays
as multi-sigma deviations (we use 3 standard deviations in a primary
search step, then allow ``growth'' of cosmic rays by lowering the
threshold to 1.8 $\sigma$ for pixels within a radius of one pixel) of
individual data points from this image model.

This image model then serves multiple purposes: (1) allows full
sensitivity to cosmic ray detection in under-sampled and dithered
data, (2) supports image registration determination (\S 3.2), (3)
supports scattered light removal (\S 3.3), (4) and can be used to
create an over-sampled image simply by evaluating Eq. (1) at a number
of regular sub-pixel coordinates (but see \S 3.5, a different approach
is adopted here for generating a combined image in support of the SN
search).

The image model is developed by fitting the polynomial representation
of Eq. 1 to a stack of pixel values corresponding to the separately
known $x,\,y$ offsets with weights given as the inverse variance
expected from Poisson noise on the object and sky plus readout noise.
For pixels on the inner wings of bright stars the full 10 parameter
representation is required for a good match of model and data.  In
regions dominated only by sky background such a multi-parameter fit is
inappropriate and over-fits the data, producing an excessively noisy
background if used to produce an over-sampled image.  To avoid the
latter problem the number of terms allowed in the image model fit for
any pixel is adjusted as a function of mean signal level.  Also, given
the relatively small number of sub-pixel dithers, coupled with
geometric distortion some pixels will have good phase-space coverage
in the offsets and others will have a very unbalanced distribution of
sub-pixel $x,\,y$ sampling -- this is also taken into account in
determining how many terms are used to represent Eq. 1 for any pixel.
The net result of this is a {\em variable resolution} model of the
spatial distribution of light on the sky: high resolution near bright
objects, low in regions of low signal (or inherently poor dither
sampling at the sub-pixel level).  The variable resolution is
appropriate for supporting cosmic ray detection and image
registration, but quite inappropriate for a final mean image for which
underlying PSF stability is needed in support of object detection and
characterization.

\subsection{Image Registration.}

Since we wish to derive combined images with over-sampling to
optimally recover critical sampling, accurate knowledge of
frame-to-frame offsets is essential.  Fortunately, even in the sparse
HDF and low-signal F300W images there is adequate information in
compact objects to support such relative offset determination.

Geometric distortion exists in the WFPC2 cameras at a sufficient level
that it must be directly accounted for in developing the offsets for
any pixel.  We have used a simple $r^3$ distortion term with separate
centers and coefficients determined for each CCD camera as necessary
to bring images at the edges of adjacent chips (provided by the K-spot
calibration images) into registration.  Inspection of distortion maps
as provided in \cite{holtz_95} shows that such a simple radial term is
appropriate and our three parameter ($x,\,y$ center for each chip and
coefficient on the cubic term) representation provides accurate
results and is much easier to work with than the twenty-term,
two-dimensional polynomials.  For these analyses we correct for
geometric distortion only in a differential sense, our final combined
images retain the intrinsic WFPC2 distortion, but the effect of
geometric distortion in producing variable offsets as a function of
position over the CCD in response to large dithers is fully accounted
for.

A further complication in both the original and second epoch data are
small rotations (up to 0.0025 radians) sufficient to generate relative
shears of one pixel in the CCD corners if not accounted for.  Across
epochs we also cannot assume that the plate scale remains constant to
the accuracy we need ($\lesssim$0.1 pixel side-to-side on a CCD).

The image registration therefore accounts for: (1) $x,\,y$ offsets at
chip center are evaluated for every frame.  (2) Geometric distortion
is modeled to yield correct sub-pixel offsets for all individual
pixels as a function of their position in CCD coordinates and the
primary offsets.  (3) A simple rotation matrix adjustment of $x,\,y$
offsets away from chip center.  The rotation may be assumed constant
for all of the second epoch relative to the first, constant at a
different value for the four images affected within the first epoch,
and the values derived from the high S/N F814W data may be assumed for
the F300W images.  (4) A separate $x$ and $y$ plate scale for each CCD
camera to be evaluated only as the second epoch relative to the first.
A formal solution yielded a plate scale difference of $\sim 8 \times
10^{-5}$ in the $x$ direction (all chips rotated to align with WF4),
and half this in the $y$ direction across the epochs.  This yields a
relative shift of 0.06 pixels side-to-side over the CCDs.  Independent
determinations for the three WF chips showed a scatter of $\pm$ 15\%.
In generating combined images for the second epoch we have accounted
for this marginally significant, albeit relatively unimportant term.

The image model provided by Eq. 1 consists of CCD-sized arrays for
each of the 10 polynomial coefficients.  As previously discussed only
a subset of the coefficients are used (down to just the zero point in
regions of sky background) based on local data characteristics.
Furthermore during the solution for Eq. 1 on each pixel, terms are
only retained if judged significant (magnitude greater than twice the
error).  The pixel-by-pixel solution is iteratively developed
retaining only significant terms (e.g., on the edge of a bright star
all 10 terms would likely be kept, on sky only the zero point, and in
the center of a bright, but broad galaxy only the zero point).  The
image model therefore very directly reflects which pixels (those with
both a substantial signal level and several populated higher order
terms) carry significant positional information.  For a given CCD we
utilize the $\sim$1\% of pixels carrying almost 100\% of the
positional information in evaluating registration offsets.

We compute image registration offsets (and sometimes rotations and/or
plate scale changes) for individual frames by evaluating what offsets
yield the optimal (in a weighted least-squares sense) match of the
evaluated image model (Eq. 1) ``shifted'' to register with individual
frames.  Only the relatively noiseless model image is evaluated with
different shifts and this process does not involve interpolation of
individual under-sampled images.  After the first iteration of this
process the individual images have been cleaned of cosmic rays (pixels
flagged as having been hit by a cosmic ray have had data values
replaced by the image model value).  Resulting image registrations for
individual F814W images are good to $\sim$ 0.02 pixels for the basic
$x,\,y$ offsets and to $\lesssim$ 0.05 pixels side-to-side for
rotation and/or plate scale change effects.

The above approach works very well when applied to individual frames
in the original epoch in reference to the model image of the original.
Since we will search for SNe in difference images across the two
epochs, and thus need excellent registration of the two, we also
register individual frames of the second epoch against the model image
from the first.  In our first full solution it became immediately
obvious that the two epoch difference image showed poor subtractions
of both stars and galaxies.  Near the position of bright stars the
difference images tended to show negative to positive intensity
gradients in random directions -- stars move; proper motion over two
years was sufficient to generate significant registration differences
for stellar sources.  We then dropped all stars detected by Flynn
\etal 1996 from the cross-epoch registration solution and immediately
obtained excellent subtractions on galaxies.  (Stars fainter than the
Flynn sample are at too low a signal to noise to influence the
registrations.)

\subsection{Scattered Light Removal}

The WFPC2 cameras generate characteristic (but variable according to
details of illumination) large scale ``X'' patterns of excess
background light from bright Earth limb photons directed into the {\em
HST} optical truss at shallow angles.  Since added light carries a
Poisson noise penalty, even if the pattern can be subtracted out, the
best approach to scattered light is to avoid observing close to the
bright Earth limb.  However, with the availability of Continuous
Viewing Zone observations and wanting to use as much of each orbit as
possible, it is likely that a subset of the images will show low
levels of scattered light.

We define and correct for scattered light via the following approach:
(1) Using a subset of exposures entirely free of scattered light
define a model image as in \S 3.1.  (2) Use the scattered light free
image model and subtract it on a frame-by-frame basis from all images
with possible scattered light contamination.  This is done on images
that have been cleaned of cosmic rays.  (3) Smooth the resulting
images using Fourier filtering and subtract the resulting images from
the individual data arrays before performing a final combination.
This approach works well and provides a final combined image with no
trace of the scattered light pattern.

\subsection{Delta-Dark and Hot Pixel Definition}

Scaled to a typical exposure time (2700s) of individual input frames
the final combined HDF F814W images should have an rms noise in
regions of sky of $\sim$1.85 e- and 2.65 e- for the first and second
epochs respectively, or 3.23 e- for the difference image.  Any ``hot''
pixels with count rates $\gtrsim$3.23/2700 = 0.0012 e-/s, or on
average less than 1 DN (at a gain of 7 e-/DN) per integration will
thus contribute equally to the final noise (or start to rise above it
and thus mimic an event).  The darks can be determined to a level
several times better than this ($\sim$2.2 $\times 10^{-4}$ e-/s rms
first epoch, and $\sim$3.8 $\times 10^{-4}$ e-/s for the second epoch;
for the longest exposure times in both epochs this corresponds to 0.7
e- rms uncertainty in the dark contribution for typical pixels per
readout) as we argue next.

The HDF data using multiple available bandpasses and augmented with
any contemporaneous real dark exposures can be used very effectively
to define a delta-dark (a primary dark image has been removed already
in the pipeline reductions for each frame) and to search for any hot
pixels, including especially those that only become hot during the
interval of the observations.  Using image models as defined in \S 3.1
we create individual frames that have been: (1) through standard
pipeline reductions, (2) cleaned of cosmic rays, but with retention of
cosmic ray images (i.e., the spatial distribution of light attributed
to cosmic rays on a frame-by-frame basis), (3) cleaned of any
scattered light, and (4) subtracted using the image model (i.e., the
real sources are removed).  The resulting images should simply be
random noise (sky and object Poisson plus readout), plus any dark not
yet accounted for.  These resulting images are all scaled to a common
exposure time and stacked in direct pixel space.  (To search for hot
pixels with large amplitude the cosmic rays must be added back in,
since at least in the second epoch data the frequent dithering would
result in their elimination as cosmic rays.)  A consistently hot pixel
will stand out as having a positive value throughout and fluctuations
as expected from Poisson and readout noise.  For each pixel a sample
mean is formed for values within the 10th and 90th percentiles using
appropriate weights from each individual pixel-frame (a pixel that was
sitting on a bright target in a given frame would have a low weight
due to Poisson noise).  This becomes a new delta-dark image.  All
pixel time series are checked for a number of anomalies and if any of
the following tests ``succeed'' then the pixel is flagged as bad and
not used in forming a final combined image: (1) absolute value of the
dark rate exceeds 0.008 e-/s, (2) the standard deviation on the pixel
time series is more than 1.5 times the expected value, (3) the
absolute value of difference between the first 10\% (in time) and last
10\% ratioed to the overall standard deviation exceeds 3.5 (these are
pixels that turned on or off during the observing period), (4) a
linear correlation coefficient between pixel values and local sky
level exceeds 0.25 (evidence of a low amplitude non-linearity or
trap).

The second epoch data used to define darks included 76 F814W, F300W
and dark images totaling a dark accumulation time (slightly longer
than exposure time) of 122,160 s on WF2.  The original epoch data used
included 157 F814W, F300W and F450W images totaling 374,100 s. (Darks
had not been reduced using the same master dark as the data frames;
since the available data was more than sufficient already the first
epoch dark frames were not brought in.  Images with either short
exposure times, or excessive sky levels were dropped.)

Summed over the four CCDs 2222 pixels were already flagged as bad by
the pipeline (we keep these flagged as bad), after augmentation by the
above tests we carry an additional 15,465 bad pixels (second epoch,
similar size, but different list in first) -- still only 0.8\% of all
pixels in active imaging areas.  The remaining $>$99\% of pixels with
delta-dark rates $\leq$0.008 e-/s are then additionally corrected for
these (appropriately scaled dark integration time) delta-dark values.
Our combined images (next section) have marginally better (averages
2.1\% for the three WF chips in F814W) levels of background noise in
comparison with the Version 2 HDF release (Williams \etal 1996) and
significantly better at 21.8\% for the single case (PC1 and F300W)
most sensitive to having details of the dark current and elimination
of slightly bad pixels done optimally.

We have been particularly diligent in defining (and then ignoring) hot
pixels, since for SN detection in undersampled data a hot pixel at
$\sim$0.035 e-/s provides the same count level as a SN of $m_I \,\sim$
27.3 near our detection limit.

\subsection{Image Combination}

We have already described in \S 3.1 an image model that provides one
realization of image combination algorithms.  However, its
over-fitting of noise in regions of low signal, or resolution
dependence on local signal if the fit order is varied makes this a
poor realization in which to search for faint point sources.  (This
technique is ideal for use as an intermediate tool for eliminating
cosmic rays, deriving image registrations, etc.)

In support of SN detection a combined image for each epoch that
retains all available spatial information, maintains photometric
stability and reaches the noise floor specified by simple Poisson
statistics and readout noise is desired.  A strong candidate for
providing this was the ``drizzle'' software used in the original
release of combined HDF images.  A minor weakness noted in Williams
\etal 1996 for the drizzle approach was that PSFs on the brighter
stars showed ``noticeable high-frequency noise'' (seen even in
simulations without Poisson noise).  We have chosen a more
``continuous'' approach to assigning values on an over-sampled grid,
than the discrete apportionment of flux from individual WF pixels used
by drizzle; in principle this should produce a more stable PSF.  In
practice our approach does seem to provide a slightly sharper and
smoother PSF than was realized in the Version 2 drizzle results,
although real gains are marginal at best at a level of a few percent.

Our algorithm is even simpler than drizzle and consists of the
following: (1) Start with a stack of frames for which all available
corrections (calibration reductions for bias, flat-field and dark; sky
and scattered light subtracted; a delta-dark subtracted) have been
made and for which frame-to-frame registration is accurately known as
well as knowledge of all pixels affected by cosmic rays.  (2) Define a
regular grid of over-sampled positions in pixel space (or on the sky
if geometric distortion removal in this step is desired) upon which
the combined image will be defined.  We use a regular grid of
$\times$4 over-sampled sub-pixels, the resulting image retains
geometric distortion.  (3) Adopt a weighting function to transfer
values from pixels with arbitrary, nearby offsets to an accumulation
at the grid positions.

We use a simple Gaussian weighting with a width parameter of $\sim$0.3
WF pixels.  The intensity at an arbitrary $x,\,y$ grid position is:

\begin{equation}
I_{xy}\;=\;\frac{ \sum_{i=1}^I \sum_{j=1}^J \sum_{n=1}^N gw_{i,j,n}
pw_{i,j,n} I_{i,j,n} } { \sum_{i=1}^I \sum_{j=1}^J \sum_{n=1}^N
gw_{i,j,n} pw_{i,j,n} }
\end{equation}

\noindent
where the Gaussian weighting is specified as:

\begin{equation}
gw_{i,j,n}\;=\; e^ - \left(\frac { (x_{i,j,n} - x)^2 \,+\, (y_{i,j,n}
- y)^2 }{ \sigma ^2 } \right)
\end{equation}

\noindent
and the $pw_{i,j,n}$ are pixel weights set as the inverse variance sum
of Poisson and readout noise components.  The $x_{i,j,n}$ and
$y_{i,j,n}$ are the relative offsets of a given pixel accounting for
geometric distortion, the frame-to-reference image registration
offsets, rotation and plate scale changes.  The range of $n$ is over
the full stack of images (58 F814W in first epoch, 36 for second), and
in principle $i$ and $j$ range over the full indices on each frame,
but the Gaussian cut off is steep enough that in practice only pixels
within $\pm$1 pixel index (after allowing for multi-pixel dithers to
nearest integer) are summed over.  The value of $\sigma$ is
adjustable; small values result in the best resolution possible, large
values in compromised resolution but lower background noise
fluctuations.  We selected $\sigma$ = 0.3 by running a number of
trials to optimize S/N for faint point sources taking into account the
actual resulting PSF widths and background noise levels in combined
images.

For each combined image an error array is carried that reflects the
amount and quality of data going into any grid point accumulation:

\begin{equation}
Err_{xy}\;=\; \left[ \sum_{i=1}^I \sum_{j=1}^J \sum_{n=1}^N gw_{i,j,n}
pw_{i,j,n} \right] ^{-1/2}
\end{equation}

Our resulting PSF widths and background noise levels for the original
epoch data are very similar to, but if anything marginally better
(clearly better for the PC and F300W where our improved treatment of
darks was probably significant) than the Version 2 drizzle release
(Williams \etal 1996).  Table~\ref{tab2} provides a quantitative
comparison of results.  Properties of the second epoch combined image
were as expected: resolution the same as for the original epoch, and
higher noise resulting from a factor of two less integration time in
F814W.

\placetable{tab2}

Only for the PC1 F300W image combination are differences discernible
between the two reductions of the first epoch HDF data from a careful
visual inspection.

Image combinations for the second epoch were carefully compared with
the first epoch equivalents for PSF widths, registration and noise
level.  No remaining systematic effects were found (except for the
expected higher noise level in the second epoch).  For some individual
stars there are noticeable PSF differences between epochs, probably
following from the rather meager sub-pixel $x,\,y$ phase coverage
provided by limited dithering for the first epoch HDF data.  These
differences are, however, small and not likely to cause problems for
SN searches in differences of first and second epoch combined images
(and the minor differences are accounted for in control experiments
discussed in \S 4.3).

\subsection{Point Source Photometry}

The final product of image combinations described in \S 3.5 are images
still geometrically distorted as in the original data, but now
$\times$4 over-sampled, and scaled to the number of e-/pixel detected
in a 6000 s exposure.  The photometric transforms and basic integrity
of these combined images were determined by: (1) The intermediate data
products consisting of individual images scaled to e-/2700 s with
corrections made for sky, scattered light, cosmic rays, and
delta-darks were transformed back to DN/s using the same gain assumed
in initial transforms to e-.  This provides data frames that can be
analyzed using standard prescriptions for WFPC2 photometry (e.g.,
WFPC2 Photometry Cookbook and WFPC2 DATA Analysis -- A Tutorial, both
available under the ST ScI Instruments WWW pages).  (2) We selected
the 11 Flynn \etal (1996) stars on WF2, WF3 and WF4 with magnitudes
between $m_I$ = 19.0 -- 25.5; these stars never saturate in the F814W
images and have adequate per exposure signal on the fainter stars.
(3) Using {\em all} of the photometric corrections described in WFPC2
photometry documentation from ST ScI (0\farcs5 radius apertures,
geometric distortion correction, CTE including time dependence, minor
chip-to-chip normalizations, and transform to standard Johnson-Cousins
\I for F814W) magnitudes were computed for these 11 bright stars.  (4)
The same 11 stars are used to establish a photometric transform by
adopting the magnitudes resulting from step (3) and measuring
intensities in the over-sampled images using the same DAOPHOT
parameters used in the SN searches to be detailed in \S 4.2.

Results from these photometric tests may be summarized as: (1) From
the original epoch data our $m_I$ magnitudes have a zero point shift
of only 0.023 (sign of ours - Flynn) and an rms scatter of 0.05
magnitudes relative to Flynn \etal (1996).  (2) Across the two epochs
our measures show a zero point offset and scatter of 0.015 magnitude.
(3) Within cameras, transforming the over-sampled image photometry
using DAOPHOT PSF fitting with a simple zero point offset relative to
the known single-frame, aperture photometry derived values yields a
scatter of $\sim$4\%.  We expect photometry errors on our SN
detections to be dominated by random rather than any residual
systematic errors.

For ease in comparing these results to other publications we note that
the following transforms from our standard Johnson/Bessel-Cousins
photometry apply:

\begin{eqnarray}
m_{I, Vega-mag} = m_I - 0.035 \\
m_{I, AB(8000)} = m_I + 0.396 \\
m_{I, ST(8000)} = m_I + 1.220 
\end{eqnarray}

\section{Supernovae Detection and Significance}

The previous section outlined the steps taken to produce a single
combined F814W image for each of the WFPC2 CCDs in both the original
and second epoch HDF observations.  The two epoch images
(3200$\times$3200 arrays for each CCD) are co-aligned to a small
fraction of a pixel using extra-galactic sources and are normalized in
intensity.  A simple difference image therefore yields excellent
cancellation for stationary, non-variable objects; those with changed
position or brightness show intensity gradients or localized net
intensity changes respectively.  In this paper we are interested in
detecting SNe and therefore search for point sources present in one
epoch, but not the other.  In addition the image combination process
provided an error array; the image ratio of difference to error arrays
provides a simple means of assessing the significance of intensity
changes across epochs.

\subsection{Inspection of Difference Images}

Both of the SNe detected in the HDF were first found via a rather
casual inspection of direct, and in ratio to the error array,
difference images.  These two events remained the only convincing
candidates following detailed systematic inspection of the difference
images by both RLG and MMP.  Figure~\ref{fig1} shows a small region
around SN 1997ff in the first and second epoch direct images and the
difference: second minus first.  Figure~\ref{fig2} shows the same
(both events were by chance in the second epoch) for SN 1997fg.  A
slightly broader perspective is shown for both events in
Figure~\ref{fig3} which provides some qualitative indication (the two
panels combined cover $\sim$2.4\% of the area surveyed on the WF CCDs)
of the general quality of galaxy subtractions in these data.  The
stable PSF provided by {\em HST}, even across a time span of two years
supports excellent cancellation of non-variable objects.

\placefigure{fig1}

\placefigure{fig2}

\placefigure{fig3}

Before providing quantitative details on SN 1997ff and SN 1997fg
(previously announced in \cite{hdf_iau_98}) we will now consider in
turn a quantitative search for SN candidates in both epochs, null
experiments designed to indicate at what level false signals may
arise, and a quantification via artificial star additions to the real
data of completeness as a function of object brightness and host
galaxy redshift.

\subsection{Quantitative Search for SN Candidates}

We have used the DAOFIND, PHOT and NSTAR routines in DAOPHOT
(\cite{stetson_87}) with parameters set appropriate to the PSF scale
(FWHM of PSF = 5.7 pixels in our $\times$4 over-sampled data) and
general background noise level ($\sim$10 e- rms pixel-to-pixel for an
over-sampled difference image scaled to 6000 s exposure time).  A
threshold of 6 $\sigma$ is used in DAOFIND which results in detection
of ``sources'' well beyond the level at which real events can be
reliably found.  The candidate SN lists are then reduced using cuts in
magnitude, magnitude error and the DAOPHOT sharpness statistic.  The
appropriate cuts were selected to maximize completeness in recovering
artificially added stars, while at the same time allowing no sources
to be claimed in control experiments designed to mimic potential noise
sources, but for which no variation of sources across epochs was
included.

A separate PSF was developed for each of the WF CCD cameras using 2-3
isolated, bright, but never saturated stars.  Although WFPC2 shows
small PSF spatial variance, we did not attempt to account for this; at
the relatively low signal to noise of candidate SN the assumption of a
spatially invariant PSF is warranted.  Both the first minus second and
opposite difference images are independently searched since DAOPHOT is
structured to only find positive stars.

A primary cut in candidate SNe will be made using the magnitude error
provided by NSTAR in DAOPHOT.  There are two factors in addition to
the mean background fluctuation level that need to be considered: (1)
local errors will be larger in the difference image if an underlying
object provided extra Poisson noise, or if an imperfect registration
led to imperfect cancellation, and (2) the pixel-to-pixel errors are
highly correlated in an over-sampled image.

It is instructive to consider at what level an increase of effective
noise in the difference image arises from underlying sources.  The sky
rate in F814W averages 0.0446 e-/s per WF pixel, over the sum of both
epochs this yields an average of 8320 e-/pixel to which must be added
equivalent readout noise of some 2820 e-/pixel from the 94 separate
exposures.  Therefore an underlying object will raise the errors by
$2^{1/2}$ at a count rate per pixel of $\sim$0.06 e-/s, which for the
central pixel of a point source occurs at $m_I \,\sim$ 25.5 (since our
suspected SNe are well under this they do not significantly perturb
the per pixel errors).  This same count rate corresponds to an object
with a surface brightness of $m_I \,\sim$ 21.8 magnitude per square
arcsec, which for a galaxy with half-light radius of 0\farcs3
corresponds to a fairly bright (by HDF standards) 24th magnitude
galaxy.  Most of the surface area in the HDF is sky + readout noise
dominated.  Nonetheless, on bright galaxies there is added noise and
we account for this by scaling the DAOPHOT reported magnitude errors
with the PSF-weighted local value of the error array in ratio to the
sky dominated value.

We account for the pixel-to-pixel noise correlation by renormalizing
the DAOPHOT provided magnitude errors to average in the mean the
standard deviation in magnitude of a large sample of added bright
artificial stars (the required multiplicative factor is 2.0 for these
$\times$4 over-sampled data).  The difference images show very little
variation in mean, residual background level; we found that
significantly smaller scatter in derived magnitudes of artificial
stars resulted from forcing DAOPHOT to assume a constant (zero) sky
level.  Since the combined images have also been cleaned of cosmic
rays, parameters in DAOPHOT were adjusted to not allow a dynamic
re-adjustment of pixel weights normally used to reject cosmic rays
during PSF fits.

\subsection{Null Experiments (Controls)}

Since we expected only order one to a few events in this project it is
essential that a means be developed of quantifying the detection limit
beyond which false events would begin to enter.  We have designed two
sets of control experiments with noise characteristics very similar to
the real two-epoch image differences, but for which no true sources
can exist.

For the first of these null experiments we produced combined frames using
the data from both epochs, but including in one set only the even numbered
frames and in the other set the odd numbered frames.  Difference images were then
formed as usual between these two ``epochs''.  Any event with a
timescale long compared to single exposures of $\sim$2700 s should
thus be canceled out.  This uses the real data so noise
characteristics must be the same overall, except in the real two epoch
difference the exposure times are not balanced, and also any residual
large-scale, systematic registration errors would enter differently
than for the even-odd case.  Another caveat arises if the exposure
times in either epoch varied in an even-odd sense; in fact in the
second epoch 1600 and 1900 second exposures were alternated so real
epoch differences would survive, but at a reduced level of $\sim$
(300/1750)/3 (at this level neither of the real SNe are detectable in the
even-odd difference image).  As with the real data, both the even-odd
and odd-even difference images are analyzed with the DAOPHOT codes
using the same control parameters as for the real data.  Since such a
test includes most, but not necessarily all sources of noise which
could affect the real epoch differences, it is a necessary, but not
sufficient condition to believe the claimed SN detections that no
events be detected that are as significant in terms of the selection
criteria of magnitude, magnitude error, and the sharpness parameter.
With more relaxed selection criteria than will be used for final SN
claims the most significant false signal from this even-odd control
had $m_I$ = 27.86, mag-err = 0.20, and sharpness = -0.19 -- the
magnitude is significantly fainter and the associated error larger
than those for the two SNe found by inspection (and of course the
quantitative DAOPHOT analyses).

Furthermore, this event was found near a CCD edge ($>$90\% out from
the center in $y$) and superimposed on a bright ($m_I$ = 21.4), z =
0.562 galaxy.  A minor defect exists in the scattered light
subtraction that produces a small amount of extra noise
(frame-to-frame so it would be present in this even-odd control) at
the location of bright objects at a CCD edge (one aspect in which our
reductions are marginally poorer than the Version 2 drizzled images
discussed earlier).  We have now allowed for this (or other
unrecognized systematics contributing to noise in the control
experiments) by further biasing up the magnitude error by the ratio
(if larger) of local noise in the control cases to the equivalent in
the error array.  This is a generally small effect, but properly
accounts for lower confidence in regions with imperfect systematic
reductions.  This adjustment in magnitude errors appears throughout
this paper.

The second type of control experiment is a direct attempt to simulate
the data noise characteristics.  The image model (using that from the
first epoch only since we want the two simulated epochs to be
identical except for independently added noise) is adopted as a
noise-free representation of the image (true in comparison to single
exposures), these are then used to generate a simulated data frame
with Poisson and readout noise added on at each individual offset
image position.  These single images are then reanalyzed as with the
real data frames (except the delta-dark was not added in and thus
subtracted out) to produce combined first and second epoch images, the
resulting noise characteristics do accurately reproduce the real
difference images.  Again it is a necessary, but not sufficient
condition to believe the reality of the claimed SNe that this control
experiment not show any events as significant.  In this case the most
significant event had $m_I$ = 27.85, mag-err = 0.37 and sharpness =
-0.07; again the magnitude is significantly fainter than for the
claimed SNe and the mag-err much larger.  This event happens to fall
on an $m_I$ = 20.9, z = 0.764 galaxy, not particularly close to the
field edge.

The top 5 (in magnitude order) false alarm events (all detections from
the control experiments with $m_I$ $<$ 27.9 and magnitude error $<$
0.4) are shown in Table~\ref{tab3}.  We must select criteria that will
exclude all of the false signals found in the forward and backward
differences for both the even-odd and simulated data control
experiments, but will first examine the more significant events found
in the real data differences.

\placetable{tab3}

Table~\ref{tab4} shows the brightest 7 potential SNe detected in both
epochs (all detections from the real data with $m_I$ $<$ 27.7 and
magnitude error $<$ 0.3).  The candidate SNe in lines 3, 5, 6 and 7 of
Table~\ref{tab4} are each clearly associated with diffraction spikes
(at $\sim$1\arcsec distance) on the two brightest HDF stars that did
not cancel out perfectly between the two epochs.  No masking was
applied near the bright stars in the DAOPHOT based search; the stable
images provided by {\em HST} led to clean epoch-to-epoch differences
with few spurious events that cannot simply be attributed to random
noise.  The two claimed SNe and the potential event on WF3 from line 4
of this table will be discussed in detail below.

\placetable{tab4}

\subsection{Artificial Star Experiments (Completeness)}

In order to assess our sensitivity to SN detection we have generated a
large number of artificial star experiments adding in ``SNe'' of known
brightness to the difference images (and with both signs as
appropriate for first and second epoch events).

To avoid counting potentially spurious events that might arise from
noise, thresholds have been set as $m_I \: <$ 27.7, mag-err $<$ 0.2,
and $|$sharpness$\times$mag-err$|$ $<$ 0.06.  These cuts eliminate
(with considerable margin) all of the control experiment events listed
in Table~\ref{tab3} as well as the only viable SN candidate (in
addition to the two firm detections) listed in Table~\ref{tab4}.  In
practice the simple limit on apparent magnitude is the most
significant factor in defining the final sample.  We have tested for
completeness for events that might come from very faint galaxies not
visible even in the HDF by adding artificial stars at random positions
over the field.  In order to account for changes to noise
characteristics due to galaxies (either through added Poisson noise,
or through imperfect registration and subtraction) we have added
artificial stars to galaxies identified in the \cite{soto98} table and
tracked completeness independently for the galaxy redshift ranges 0.0
$<$ z $<$ 0.4, 0.5 $<$ z $<$ 0.9, 1.0 $<$ z $<$ 1.4, and 1.5 $<$ z $<$
1.9.  Typical galaxy magnitudes are: z = 0.2, $m_I \;\sim$21; z = 0.7,
$m_I \;\sim$25; z = 1.2, $m_I \;\sim$26, and z = 1.7, $m_I \;\sim$27
with the FWHM of magnitude histograms being about 3 magnitudes around
the above medians.  Galactic first moment radii have been adopted from
Williams \etal (1996).  We note that for a deVaucoulers' $r^{1/4}$ law
for galaxy brightness distribution the half-light radius is one-half
the first moment radius.  The artificial SNe are added to the
difference images with a random Gaussian position distribution such
that each has a 50\% chance of falling inside the half-light radius.
We have also preferentially added the artificial SNe with a frequency
proportional to the galaxies' intrinsic brightness within each
redshift bin.  The number of added SNe (both epochs) totaled 64, 322,
262 and 230 for the z $\sim$ 0.2, 0.7, 1.2 and 1.7 redshift bins
respectively.  The random positions were set once for each galaxy
redshift range and then held fixed for independent additions at the
different magnitude levels.

Table~\ref{tab5} shows the completeness limits derived from these
artificial SN experiments.  The results for both epochs have been
averaged with no difference having been noted between the two epochs.
The completeness level for artificial SNe added into blank regions of
the HDF is well represented by the limits for the z = 1.7
galaxy-redshift bin; at this galaxy brightness the difference image
noise level is set by sky plus readout noise and is little influenced
by an underlying faint galaxy.  While $m_I \;\sim$ 27.5 (at high z) is
about the 50\% completeness level {\em for which we are also confident
that objects brighter than this have little chance of having arisen
from noise}, the actual detection levels with DAOFIND are about
one-half magnitude deeper.  For example at $m_I$ = 27.9 the DAOFIND
search phase returns 30\% and 40\% of the added artificial SNe at z =
0.7 and 1.7 respectively, but since our magnitude cut is well above
this only the few having extracted magnitudes too bright by $>$0.2
magnitude and with errors $<$ 0.2 magnitudes are counted in
Table~\ref{tab5}.

\placetable{tab5}

\subsection{SN 1997ff, SN 1997fg}

We have shown the direct and difference images for SN 1997ff and SN
1997fg in Figures~\ref{fig1} -- \ref{fig3} and in previous sections
presented arguments related to significance of the detections.  In
this section we further develop the case for the events being SNe in
host galaxies at high redshift.

A SN candidate should pass several tests: (1) Signal level well above
background fluctuations and above the level at which false alarms in
similar searches would begin to show up.  (2) The point spread
function should have a distribution similar to that of other stellar
sources in the field.  (3) The objects should be detectable in
independent subsets of the data including divisions in which different
dither positions on the CCD were used.  (4) The object intensity
should show changes (or near constancy) over the multi-day observing
interval consistent with expectations for a SN light curve transformed
to the appropriate redshift (intensity adjusted including appropriate
K corrections, and dilated in time).  (5) If located in an obvious
galaxy the SN candidate should be positioned such that active galactic
nuclei (AGN) fluctuations are not an equally, or more likely
explanation of the intensity change.

We have argued above that the two SN candidates satisfy the first test
-- the quantitative magnitudes and associated errors returned with
DAOPHOT processing are significant when compared to our sensitivity to
spurious events arising from noise.  A simpler approach of placing
small apertures on the SN candidates and ratioing the aperture counts
to the error from enclosed background fluctuations shows significance
near 20 and 9 $\sigma$ for SN 1997fg and SN 1997ff respectively.

Figure~\ref{fig4} shows radial profiles for the two SN candidates from
difference images and similar profiles for nearby stars from the
second epoch direct images.  For each of the SN the profiles are
stellar in appearance and Gaussian fit widths for the SNe are the same
to within expected errors (i.e., fit would not be significantly worse
if the width were forced to the stellar value) as for bright stars.
(Note also that a Gaussian fit is a rather poor approximation in
detail to the stellar profiles underestimating both the inner core and
the wings.)

\placefigure{fig4}

The second epoch observations were taken in three separate {\em HST}
visits each of six orbits with a total separation of three days.  The
two claimed SNe are apparent from inspection of each of these three
separate averages, each of which had an independent set of dithers.
The claimed SNe events clearly do not arise from isolated noise
events.  The individual daily magnitudes are given in Table~\ref{tab6}
for the two SN.

We adopted a different strategy than DAOPHOT analysis of the
over-sampled, combined data for analysis of the time-resolved
magnitudes.  Photometric precision, or integrity of the combined
images relies on having a dense sub-pixel dither distribution on the
spatial scale of the weight kernel of Eq. 3 -- this is well satisfied
for the full epoch stacks, but starts to break down for the daily
averages.  To account for this we adopted the approach of using
DAOPHOT to generate an over-sampled PSF at the position of the SN as
determined from the full epoch fit.  We then used the individual
time-series data frames (with cosmic rays eliminated and all other
corrections made) in conjunction with known dithers for each frame to
place each pixel center in a 4$\times$4 pixel region near each SN
relative to the PSF from which an appropriate estimate of the PSF was
interpolated.  In this approach a full least-squares point spread
function fit was determined as a direct summation over individual
pixels in the stack.  Since this allowed full use of positional
information, precision of the final results was not dependent upon
having a dense sub-pixel distribution of under-lying dithers.  A
separate zero point calibration for each day and CCD was determined by
analyzing the known bright stars.  Analysis of the difference image
data as required for the SNe additionally required subtracting a first
epoch intensity value determined for each $x$, $y$ position from
interpolation in the over-sampled, complimentary epoch image.  The
mean magnitudes given in Table~\ref{tab6} follow from the same
procedure, but over the full data stack, and are thus independent from
the Table~\ref{tab4} results from DAOPHOT.  Using DAOPHOT on
single-day, combined images gave similar, but somewhat noisier
results.

\placetable{tab6}

Figure~\ref{fig5} shows the daily photometry values overplotted on
canonical SNe~Ia light curves adjusted in magnitude and time dilated
according to the host galaxy redshifts and assumed cosmology
($\Omega_M$ = 0.28, $\Omega_{\Lambda}$ = 0.72, $h = 0.633$, $M_B$ =
-19.46) and intrinsic SN~Ia peak brightness.  The date of peak
brightness has been arbitrarily shifted to have the observed values
fall on the light curve on the descending branch.  The magnitudes for
both SNe show a decline (only marginally significant) easily
consistent with the $\sim$0.1 magnitude decline expected during this
time on the declining branch.  At the same magnitude relative to peak
brightness on the rise an $\sim$0.4 magnitude brightening would be expected over
the three day interval which is inconsistent with the data.  The
observed time-resolved magnitudes are entirely consistent with
interpretation as SN~Ia slightly past peak intensity.

\placefigure{fig5}

We have adopted the \cite{soto98} redshift for 4-403.0 based on their
use of 7 color ($UBVIJHK$) photometry to develop photometry-redshift
relations; \cite{connolly_97} have also emphasized the need for
near-IR colors in determining photometric redshifts over 1 $<$ z $<$
2.  For 4-403.0 the IR ($JHK$) intensities from the work of
\cite{dickenson98} are 2.5 -- 3.5 magnitudes brighter than in the
optical.  \cite{sawicki97} based on a photometric redshift technique
using only the direct HDF $UBVI$ photometry determined a redshift of z
= 0.95 (their object 40368).  Our interpretations will assume the z =
1.32 value for SN 1997ff, although we note that without a
spectroscopic determination this must be viewed with caution.
The \cite{soto98} photometric redshifts show an $rms$ dispersion of 
$\sigma$ = 0.09 in comparison with a sample of $\sim$100 spectroscopic
redshifts at $z$ $<$ 2, and no cases were discordant at $>$3 $\sigma$.

SN 1997ff is located 0\farcs11 west and 0\farcs11 south of the center
of galaxy 4-403.0 \cite[]{williams_96}.  This galaxy is very red at
$v-i$ = 1.64 and is classified as an elliptical galaxy
on the basis of colors and template fitting by
\cite{soto98} (no. 690 in their table) with a photometrically
determined redshift of 1.32.  Close inspection of this image in a
multi-color representation of the HDF shows nothing to suggest
anything other than a normal, isolated elliptical galaxy.
\cite{williams_96} give a first moment radius of 0\farcs25 for 4-403.0
which translates to a half-light radius of 0\farcs125 assuming an
$r^{1/4}$ law, thus the SN falls near the half-light radius.  Given
the galaxy characteristics, and the clear off-center location for the
brightening, interpretation as a SN is much preferred over an AGN
brightening.  As shown in Figure~\ref{fig5} the observed brightness
for SN 1997ff is about two magnitudes below the peak luminosity
expected for a SN~Ia.

SN 1997fg is located 0\farcs31 east and 0\farcs31 south of the center
of galaxy 3-221.0.  This galaxy was assigned a neutral color of $v-i$
= 0.74 and is classified as an irregular by \cite{soto98} (no. 599)
with a spectroscopically determined redshift of 0.952
(\cite{cohen_96}).  In this case the galaxy morphology is not simple.
The claimed SN event is well offset from primary light concentrations,
the SN produced a factor of two local brightness increase relative to
the underlying galactic light at the 0\farcs1 spatial scale.  There is
no reason to suspect an AGN interpretation for this event.  At nearly
2.5 magnitudes below the peak brightness expected for a SN~Ia at z =
0.95 this event could arise from other types of SNe, in particular the
galaxy type and event brightness would be consistent with a SN~II
event.  We will provide further arguments in \S 5 to quantify the
relative probability that this was a SN~Ia or SN~II event.

\subsection{Other SN Candidates in the HDF}

To the level of our conservative magnitude and error cuts used to
quantify completeness limits ($m_I$ $<$ 27.7, error $<$ 0.2
magnitudes) there are no additional candidates once obvious artifacts
from diffraction spikes on bright stars are excluded.  The event for
the 4th line of Table~\ref{tab4}, however, deserves further
investigation as the next most likely SN candidate.  The event with
$m_I$ = 27.35 and an associated magnitude error of 0.284 is apparent
as a brightening near the core of a z = 0.52, $m_I$ = 23.60 galaxy
number 3-404.2 with a half-light radius of 0\farcs15.
Table~\ref{tab5} shows that at this magnitude and redshift artificial
SNe are only recovered at an efficiency of $\sim$25\%.  False alarms
from our control experiments fall short by 0.5 magnitudes in reaching
this level however, thus putting this event in a grey area relative to
our reasonable, but somewhat arbitrary choice of selection criteria.

The position of the 3-404.2 excess light in the first epoch is only
0\farcs048 from the galaxy center, since this galaxy has an elongated
shape and the candidate SN signal is weak the measured offset from the
nucleus is not considered significantly different from zero.  The
amount of light interior to a radius of 0\farcs05 for galaxy 3-404.2
is $\sim$4\% of the total.  These facts suggest an alternative
interpretation as an AGN is preferred.  We have performed an analysis
comparing the change in total intensity within the half-light radius
of 3-404.2 across epochs (4.9\%) with expected noise levels (1.2\%) at
this galaxy brightness from simulations, thus establishing at
$\gtrsim$4 $\sigma$ that this galaxy did exhibit a real intensity
variation -- most likely from nuclear activity, although we cannot
rule out a SN interpretation.

An additional variable source, galaxy 2-251.0, is evident from a
casual inspection of the two epoch F814W difference image.  This
bright ($m_{AB(814)}$ = 21.3), compact source only marginally resolved
at WFPC2 resolution has been widely discussed in the literature as an
AGN.  The optical spectrum yields a redshift of 0.960
(\cite{acphillips_97}) and shows broad Mg II emission and other weaker
features consistent with nuclear activity.  In the radio it displays a
core size $<$ 0\farcs1 and is variable by about 30\% over 18 months
(\cite{fomalont_97}, \cite{richards_98} ), and it is a strong IR
source detected by ISO (\cite{rowanrob_97}).  Using an aperture of
0\farcs5 we find that 2-251.0 shows an optical increase of 8\% from
the first to second epoch and that in comparison to the expected
measurement noise this is significant at $\sim$7 $\sigma$.  The
brightening comes from an unresolved source that is coincident with
the galaxy core.  If this probable AGN variation were attributed to a
SN event instead, it would have $m_I \; \sim$ 23.5, or just marginally
brighter than a Type Ia peaks at for this redshift.  The object is
detected in our DAOPHOT search, but after error scaling as discussed
in \S 4.3 has a magnitude error in excess of unity and is not
retained.  The two-epoch difference image of 2-251.0 is sharp, but not
well matched by a PSF -- a detailed examination shows that the
sub-pixel sampling of the first epoch data was unusually poor at the
position of 2-251.0 yielding low confidence in details of the
resulting difference image morphology.  In this case we cannot rule
out a SN event at $m_I \; \gtrsim$ 23.5 near the core of 2-251.0, but
given the abundant evidence consistent with an AGN interpretation have
no reason to expect such.  This example of potential incompleteness in
our SNe search is directly accounted for in the completeness limits as
tabulated in Table~\ref{tab5}.

We do not see credible evidence for additional SN events in the two
HDF epochs.  The stable $HST$ point spread function over time and the
uniquely well-dithered and deep data in both F814W epochs on the HDF
have supported accurate quantification of completeness limits.

\section{Modeling the Expected SNe Rates}

In the previous section we quantified the brightness levels at which point
sources present in only one of the two HDF epochs could be detected.  The
number of expected SNe in the HDF at a given epoch depends upon: (1) the
SN rate per unit redshift and (2) modeling of the length of time SNe at
different redshifts will remain above the established detection threshold. 
Several recent theoretical studies have addressed the expected variation
of both Type Ia and II rates as a function of redshift.  Typical of these
is the work of \cite{madau_98}, who model the SN rate by using observed
luminosity densities in the universe to imply a star formation history. 
With an assumed initial mass function this in turn implies the number of
stars more massive than 8 $M_{\odot}$ which on short time scales become
Type II SNe.  Type Ia events are modeled as following from binary
evolution in which double white dwarfs merge, or single degenerates evolve
over the Chandrasekhar mass through accretion of mass transferred from
their companion.  With parameterized delay times in the range 0.3--3~Gyr
for this evolution after White Dwarf formation from 3-8 $M_{\odot}$ stars
(see also \cite{ruiz_98}, \cite{sadat_98}, \cite{yungelson_98},
\cite{dahlen98}) this fixes the rate of Type Ia events. 

Modeling the length of time SNe at different redshifts will remain above
our detection threshold in the HDF requires knowledge of both the light
curves and spectral evolution.  Our observational understanding of SNe~Ia
has grown during the last 10 years to the point where it is now possible
to carry out relatively precise predictions of the visibility of these
events as a function of redshift.  Calculating such ``detection windows''
for SNe~II is, unfortunately, more difficult due to their extremely
heterogeneous nature. To date, only a few attempts have been made at
carrying out such calculations for either type, and these have generally
assumed black-body (or cutoff black-body) spectra for the SNe (e.g.,
\cite{miralda97,dahlen98}).  In this section we must, therefore, develop
the machinery necessary to translate observational knowledge of typical SN
light curves and spectra into realistic detection windows. 
For these simulations we have ignored Type Ib/Ic SNe, but note that future
studies with higher detection statistics should include these types as well.

\subsection{Type Ia Supernovae}

The magnitude of a SN~Ia in filter $X$ can be expressed as follows:
\begin{equation}
m_X = M_B(t,s) + K_{BX}(z,t,s,A_B) +
\mu(z,\Omega_M,\Omega_\Lambda,H_o) + A_B.
\label{kcorr}
\end{equation}
Here $t$ refers to the epoch when the SN~Ia is being observed, $s$ is
the stretch-factor (as described in \cite{perl97}), $z$ is the
redshift and $A_B$ is the reddening in the host galaxy. The extinction
in our galaxy in the direction of the HDF is negligible and thus is
not included. The stretch-factor, $s$, has three effects on the
supernova. First, it is responsible for changing the shape of the
light-curve. From an $s=1$ template, which is similar to the
Leibundgut template for SNe~Ia \cite[]{brunophd}, all known
light-curves for SNe~Ia can be reproduced quite reasonably from the
$U$-band through the $V$-band (over the time range of -18 days to +40
days with respect to maximum light) by stretching the time axis of the
light-curve about maximum light by the factor $s$. Second, a
relationship exists between the peak magnitude in $B$ for a SN~Ia and
$s$ that can be expressed as follows:
\begin{equation}
M_B = M_B(s=1) - \alpha*(s-1).
\end{equation}
This is the same relationship found by \cite{phil93} and studied by \cite{rpk95} where a
correlation between the peak brightness and the decline of the light
curve from maximum light to +15 days ($\Delta$m$_{15}(B)$) was
presented. The broader light curves (smaller $\Delta$m$_{15}(B)$) are
intrinsically brighter, and the narrower light curves (larger
$\Delta$m$_{15}(B)$) are intrinsically fainter.  From \cite{perl97},
the relationship between $\Delta$m$_{15}(B)$ and $s$ is:
\begin{equation}
\Delta{\rm m}_{15}(B) = 1.96 \pm 0.17 (s^{-1} - 1) + 1.07
\end{equation}
We adopt $M_B(s=1) = -19.46$ from \cite{sandage96}, \cite{hametal96b},
and \cite{sunetal99}, and $\alpha=1.74$ from \cite{42SNe_98}.
Finally, there is a relationship between the color of SNe~Ia and $s$
which can be expressed as follows:
\begin{equation}
B-V = -0.19*(s-1) - 0.05.
\label{color_str}
\end{equation}
This equation shows that the broader, brighter SNe are slightly bluer
as well (see also \cite{phillips99}).

The biggest challenge in determining the observed magnitude is in
performing an accurate K-correction. For this search, using the F814W
filter and going to magnitudes fainter than $I=27$, it is necessary to
calculate K-corrections out to $z=2.0$. This means knowing the flux as
a function of time for a SN~Ia down to $\approx$ 2000~\AA. As can be
seen from Eq.~\ref{kcorr} the K-correction is dependent not only on
$z$ and $t$ but on $s$ and $A_B$ as well. The affect of the former two
variables are obvious: differing redshifts and spectra (due to
temporal evolution) will change the resultant K-correction. The latter
two are secondary effects which can be thought of as ``color terms''
(the weighting of the combined spectrum and system throughput changes
for spectra with different colors brought about by extinction and/or
stretch variations). In Figure~\ref{fig6} we show the light-curves of
extinction-free, $s=1.0$ SNe~Ia between $0.25 < z < 1.75$ along with a
line which demarks a limiting detection magnitude of $I=27.3$. Note
how SNe~Ia with $z < 0.50$ are visible for over a year given this
limiting magnitude.

\placefigure{fig6}

These light curves were produced by gathering together all the spectra
of well observed SNe~Ia currently available to the authors. Especially
important to these calculations were the SNe observed in the UV by IUE
and HST \cite[]{iuesne,kir92a}. A set of standard photometric
templates were then created for a $s=1.0$ SN~Ia in $UBVRI$. The
spectra were then ``flux calibrated'' in order to reproduce the
observed magnitudes of these template light curves by both adjusting
the zero point of the flux scale and applying a slope correction to
the flux so that each spectrum would have the correct color (\cite{rpk96}) for a
particular phase. The slope correction was performed by altering the
flux using the reddening law of \cite{card89} (either making them
bluer or redder accordingly). Interpolation between these spectra from
14 days before to 45 days after maximum light was performed from 2000
to 10000~\AA . Outside of this range a simple extrapolation of the
data was used to complete the light curves.

To calculate the K-correction for a different stretch value at time
$t$ two steps are involved.  First, instead of using the spectrum at
time $t$, one uses the spectrum from the ``stretched time'', $t_s$
where $t_s = s \times t$. Second, one ``color corrects'' the spectrum
from $t_s$ according to Eq.~\ref{color_str}. The resultant spectrum is
the one used to perform the K-correction. Ideally, in order to
calculate K-corrections as a function of the stretch factor, we would
like a library of spectra covering all wavelengths and epochs of
interest and have it be representative of the full range of observable
decline rates. This is not possible with the data available to us
today. However, as will be shown in Nugent et al. (1999, in
preparation), the color of the spectrum is the most important factor
in getting the K-correction right for a SNe~Ia. By using this fact it
allows us to combine different stretch spectra into one template and
get the wavelength and phase coverage which we need to perform these
calculations. By following these procedures it is possible to produce
K-corrections to the $B$ and $V$ bands from $R$ and $I$ band data to a
$z \approx 1.0$ which are good to 0.01 mag. 
The $R$ and $I$ light curves were stretched in a similar fashion as the $UBV$
light curves.  While this is a rather crude approximation for these two 
bands (good to 0.2 magnitudes on any given day over the first 60 days of
the light curve for a given stretch supernova) it does not affect the 
results of this work strongly.  This is due to the fact that past a 
redshift of 0.3 the K-corrections to the F814W filter are done using
rest-frame spectroscopy bluer than what occurs in the restframe R-band.
At higher redshifts, when
one becomes more strongly dependent on knowledge of the UV, the
accuracy of this technique is strongly influenced by the quality of
data (in particular the flux calibration) obtained by the IUE.

Given the ability to calculate a light-curve over the necessary
redshift range, we now turn our attention to the expected rate of
SNe~Ia. The observed rates for SNe~Ia have been studied by
\cite{cap_97} for $z \lesssim$ 0.1 and at higher redshift, $z \approx$
0.5, by \cite{rate_96,rate_99}. We will use the rates from the latter
work as they are the most accurate available to date, and are based on
the CCD searches by the Supernova Cosmology Project (SCP) which are
similar in nature to the type of search performed in the HDF.  The
evolution of this rate as a function of redshift (especially for $z
\gtrsim$ 1.0), is sensitive to both the SFR and details of the
progenitor model. We will ignore this effect for now and make the
assumption that the rate of SNe~Ia per unit volume is constant as a
function of redshift and can be expressed as $2.39 \times 10^{-5}$
SNe~Ia/year/Mpc$^{3}$ at z=0.5 (restframe).  We will assume the
following cosmology: $\Omega_M$ = 0.28, $\Omega_{\Lambda}$ = 0.72, and
$h = 0.633$ as our standard.

We do not expect the volume rate of SNe~Ia to be constant between 0.0
$<$ z $<$ 1.5, none of the models predict this. However, by assuming a
constant rate per volume which is normalized at z = 0.5, we may be
either underestimating or overestimating the rate at z = 0.95 or z =
1.32 (see Fig.~1 in \cite{dahlen98} or Fig.~3 in \cite{sadat_98}) and
at some level assuming a constant rate can be viewed as the conservative approach. Finally we
note that in contrast with most theoretical studies postulating an
increasing rate of both Type~Ia and II rates to z $\sim$ 1 - 1.5 with
only a modest decline by z $\sim$ 2, \cite{kobayashi_98} argue for a
strong metallicity dependence on the mechanism allowing white dwarfs
to accrete material.  This provides a cutoff to SNe Ia for [Fe/H]
$\lesssim$ -1 and predicts a steep dropoff in the cosmic SN~Ia rate
near z $\sim$ 1.3. 

In order to compare the rates observed with a theoretical rate the
following three assumptions were made. (1) SNe~Ia have an intrinsic
stretch corrected dispersion of 0.15 magnitudes (\cite{hametal96}, \cite{rpk95}). (2) The luminosity function can be approximated using a
Gaussian distribution in stretch with a $\sigma = 0.05$ (\cite{42SNe_98}). 3) The
reddening distribution is the same as seen in \cite{hatano98}.
A Monte Carlo simulation of the HDF search
was then performed using these parameters coupled with the rate of
SNe~Ia mentioned above and the detection efficiencies described in \S
4.4.

The results are displayed in Figure~\ref{fig7}.  Out to $z\sim0.4$,
essentially all but the fastest declining SNe~Ia would be detectable
in the HDF with HST for 1+ years.  Hence, our search is complete at
these redshifts, and our simulation predicts a rapid increase in
the observed SN~Ia rate out to this redshift due simply to the
increasing volume that is being surveyed.  At redshifts beyond
$z\sim0.4$, a typical SNe~Ia ($s=1$) becomes observable for a steadily
decreasing fraction of a year, reaching 50\% at $z\sim1$ and 20\% at
$z\sim1.5$.  Hence, incompleteness becomes more and more important.
However, even ignoring incompleteness, the observed rate should begin
to flatten out or turn down (depending on the exact cosmological
parameters) due solely to geometry ($dV/dz$) and time dilation
($1/(1+z)$).  These effects combine to give the flattening and strong
turn down in Figure~\ref{fig7} which occurs between $0.5<z<1.7$.
Beyond $z\sim1.75$, the steep flux decline observed below 260~nm in
the maximum-light spectra of SNe~Ia, produces a final, rapid cutoff
below detectability.

\placefigure{fig7}

The long time period for which SNe Ia at z $\lesssim$ 1.5 are above
our detection threshold (see Figure~\ref{fig6}) has as a consequence
that most discoveries are expected for events well past maximum
brightness.  The I-band magnitude at peak and magnitude at discovery
are shown in Figure~\ref{fig8} for this simulation of the HDF search.
Note that very few SNe are discovered before maximum due to the fact
that this is the time when the light curves are evolving the fastest
in brightness.  Conversely, at epochs (and magnitudes) in the light
curve where the brightness evolution is changing the slowest (i.e.,
around maximum light and during the final exponential decline phase),
the ``magnitude at discovery'' points are observed to ``pile up''.

\placefigure{fig8}

From Figure~\ref{fig7}, we conclude that in a forward and backward
search of the HDF, $\sim$0.32 SNe~Ia should have been discovered. Due
to small number statistics, this result is completely consistent with
the observations.  Note that the expected detection frequency of SNe
is a strong function of adopted cosmology as illustrated in
Figure~\ref{fig9}.  The upper curve of Figure~\ref{fig9} reproduces
the distribution shown in Figure~\ref{fig7} for our $\Omega_{\Lambda}$
dominated cosmology which yields expected rates of $\sim$50\% and
$\sim$100\% higher than for open and flat $\Omega_{\Lambda}$ = 0.0
universes respectively.  The observed rate would therefore serve as a
strong discriminator of cosmological world models if the true rate per
unit volume were constant with lookback time as we have assumed.
However, as mentioned, the SN~Ia rate should reflect the evolution of
the SFR in the universe with a time delay which is related to the
progenitor lifetime.  Depending on details of the exact redshift
dependence of the SFR (see next section) and the typical lifetime,
$\tau$, of SN~Ia progenitors (which currently is a free parameter in
the range $0.3\lesssim\tau\lesssim3$ Gyr), increases of up to a factor
of $\sim$2--10 in the rest-frame SN~Ia rate per unit volume are
expected out to $z\sim1$ and beyond (e.g., see
\cite{dahlen98,sadat_98}).

\placefigure{fig9}

\subsection{Type II Supernovae}

There are many advantages in calculating the rates of SNe~Ia. The
light curves and spectra are homogeneous and well observed from the UV
through optical. They are among the brightest SNe having a narrow
spread in luminosity and, in general, they lack strong
extinction. Finally, the local rates are becoming fairly well
established with the myriad of CCD SNe searches currently being
conducted. All of these advantages vanish when one works with
SNe~II. They are a heterogeneous class of objects. The number of
high-quality observations is small, especially in the UV. One should
make note that while the Type II SN~1987A was by far the best observed
supernova of any type, it would almost never be seen in even the
deepest of supernovae searches, being $\approx 5$ magnitudes fainter
than the typical SN~Ia. (In the HDF search an 87A event would barely
be detectable at a $z = 0.5$). In addition, it is almost considered in
a class of its own, thus practically useless in providing information
necessary for K-corrections or light curve shapes. Lastly, the local
rates of SNe~II are poorly established due to a complete lack of
knowledge concerning the luminosity function of these objects. Even
the very deep high-redshift SNe searches have not yet made significant
inroads in understanding the rates of SNe~II. An example of this can
be seen in the work of the SCP, where over 80 SNe~Ia have been
discovered yet only 5 SNe~II have been positively identified
spectroscopically, and all with $z < 0.45$. 
Out of 95 probable SNe reported in a sample of IAU Circulars by the High-Z
team 45 are confirmed SNe Ia with precise redshifts, while only 10 are
confirmed SNe II and another 10 possible SNe II all with $z <$ 0.45.
Given all of these
obstacles, the major focus of this section will be to highlight the
techniques necessary to solve for the rates of SNe~II at
high-redshift. At present, any rate determination should be viewed
skeptically, as it is strongly dependent on several assumptions whose
basis is shaky at best.

The major problem we face concerning SNe~II is their heterogeneity. In
general SNe~II can be broken down into two classes, plateaus
(SNe~II-P) and linears (SNe~II-L) a classification which is based on
the shape of their respective light curves. Within these
sub-classifications \cite{miller_branch90} showed that the dispersion
in peak brightness in the $B$-band is more than a magnitude. These
dispersions are most likely not gaussian and are brought about by two
different factors. First, SNe~II are the result of a core collapse of
a massive star. These stars can range in mass by an order of magnitude
when they explode, resulting in very different light curves and peak
magnitudes. Second, most SNe~II suffer from host galaxy extinction due
to the fact that they are massive stars located in regions of active
or recent star formation.  As \cite{miller_branch90} pointed out in
their conclusions this fact further complicates matters since the true
dispersions could be larger or smaller due to extinction --- larger if
the brighter SNe~II in their sample preferentially suffered from
extinction, and smaller if extinction only affected the fainter
ones. Finally they noted that the true mean magnitudes would likely be
fainter due to the loss of faint SNe from the observational sample.

Heterogeneity also plays a role in the K-corrections.  While classical
SNe~II-P show generally similar spectral evolution, details such as
the time spent on the plateau portion of the light curve can vary
considerably from event to event (\cite{patat93,patat94}).  Type~II-L
SNe display even less homogeneity, and narrow emission line SNe~II
(type~IIn), which we do not even consider here since they are
relatively rare, and are even more diverse in their characteristics. 
Of the 49 SNe discovered in the Calan/Tololo survey only 4 were classified
as Type IIn (\cite{hamuyetal99}).
In addition, UV data for SNe~II (with the exception of SN~1987A) is
practically non-existent.

Given this diversity, we are forced to make a number of simplifying
assumptions in our simulations.  We start by considering a universe
which produces Type~II-P and II-L SNe only.  For each event, we
generate a rest frame absolute $B$ magnitude using the mean magnitudes
and dispersions of \cite{miller_branch90} which are uncorrected for
reddening.  Given our cosmology this translates into $M_B = -16.90$
with a $\sigma = 1.39$ for SNe~II-P and $M_B = -17.42$ with a $\sigma
= 1.32$ for SNe~II-L.  We assume that each SN is affected by the same
amount of reddening, $A_V = 0.45$ mag, which is the average value
found by \cite{skeetal94} in their study of 16 SNe~II-P and II-L.  We
also adopt mean light curve shapes for these two classes using the
templates given by \cite{cap_97}.  For the K-corrections, the best one
can hope to do is provide a correction whose uncertainty is smaller
than the dispersions mentioned above and represents in an average way
the ``typical'' K-correction for a SNe~II at a given epoch. To this
end spectrum synthesis calculations using the non-LTE code PHOENIX 9.0
\cite[]{hb971,hb972} were carried out by PEN. The spectra of SNe 1979C
\cite[]{iuesne}, 1992H and 1993W (courtesy D. Leonard) were fit at
several epochs. The resultant fits of these spectra provided a nice
match to the observed data. The major emphasis of this effort was to
model the UV as well as possible. The calculations modeled the
atmospheres by assuming a 15 to 25 M$_{\odot}$ progenitor, solar
abundances and exponential density profiles.

Many in the past have made the assumption that the spectra of SNe~II
can be modeled by a blackbody with a given effective temperature
($T_{eff}$). While this {\it might} be appropriate at very early times
it definitely begins to fail after the first $\approx$ 2 weeks. For
example, \cite{miralda97} modeled the spectrum of a SNe~II-P during
it's plateau phase by a blackbody curve with $T_{eff}=7000$. In
Figure~\ref{fig10} we present a spectrum synthesis fit to SN~1993W at
40 days along with a $T_{eff}=6000$ blackbody curve which best
represents the spectrum in the optical region ($T_{eff}=7000$ was too
hot for this particular supernova). Overlayed is the effective
coverage of the F814W filter for this supernova if it occurred at a
$z=1.0$. As one can plainly see the simple blackbody curve poorly
represents the flux of the supernova in this region. In the UV the
spectra of most supernovae are dominated by the effects of iron-peak
ionic species. The UV opacity is a product of thousands of iron-peak
spectral lines which significantly depress the flux with respect to
the continuum. In Figure~\ref{fig11} we highlight this effect by
presenting the $I$-band magnitudes as a function of redshift at day 40
(in the middle of the plateau phase) for both a $T_{eff}=6000$
blackbody curve and our spectrum synthesis calculation. Both spectra
are calibrated to have $M_B = -16.27$, an average magnitude at this
phase given our cosmology. The first thing to note is the crossing
point of these two lines. It occurs at a $z \approx 0.8$ and marks the
location where the F814W filter best matches the rest-frame $B$-band
filter. Below this redshift one is sampling the optical and IR portion
of the spectrum. One can see that a blackbody curve does a reasonable
job here with the difference in magnitudes never exceeding 0.35. Above
this redshift one samples the redshifted UV.  At a $z=1.5$ there is
already a 2.25 magnitude difference between the curves with the
blackbody curve being significantly brighter. This shows the danger in
using these simple approximations for any reasonable rate calculation
which relies on knowledge of the UV.

\placefigure{fig10}

\placefigure{fig11}

Given the library of spectra generated from our spectrum synthesis
calculations, we apply a reddening of $A_V = 0.45$ mag to each and
then scale the spectra to match the $M_B$ magnitudes generated from
the \cite{miller_branch90} distributions and the \cite{cap_97}
template light curves.  In Figure~\ref{fig12} we present, for
comparison purposes to Figure~\ref{fig6}, the resulting average
SN~II-P and SN~II-L light curves as a function of redshift. While on
average the SNe~II-L's are more than half a magnitude brighter at
maximum light, the SNe~II-P stay above a given magnitude for a longer
time due to the plateau phase. In the HDF search, given the long
baseline between reference and discovery images, they become much more
likely to detect.

\placefigure{fig12}

Given a method for calculating light curves we now turn our attention
to the rates of SNe~II. Two questions must be addressed: (1) what is
the local rate of SNe~II, and (2) how much does the SFR change over
the range in redshifts we are concerned with in the HDF search ($0.0 <
z < 2.0$). The latter question is highly relevant since it is believed
that the production of SNe~II is strongly coupled with the SFR. The
results of \cite{cap_97} show that the local rate of SNe~II (with
rather large uncertainties) is a factor of 2.5--3.0 greater than that
of the rate for SNe~Ia.  For the latter, we again assume the
\cite{rate_96,rate_99} rate (bearing in mind that this value applies
at $z\sim0.5$, and therefore may be an overestimate of the local SN~Ia
rate), which implies that locally we have $6.5 \times 10^{-5}$
SNe~II/year/Mpc$^{3}$. We will follow the work of \cite{cap_97} and
divide this number equally into SNe~II-L and SNe~II-P. We now must
consider how the evolving SFR affects this number.

In Figure~3 of \cite{madau_sfr98} we see the evolution of the
luminosity density, $\rho_{\nu}$, as a function of redshift for a flat
cosmology with $\Omega_M = 1.0$ and $h = 0.50$. Over the range $ 0.0 <
z < 2.0$ there is a factor of 10 increase in $\rho_{\nu}$ in the
UV. This corresponds to a similar increase in the SFR.  The lower
redshift data ($z < 1.0$) comes from the work of \cite{lilly_96} and
suggests a rather steep evolution in the SFR to $z=1.0$ and beyond
when combined with the work of \cite{connolly_97}. In both
\cite{gronwall_98} and in \cite{treyer_98} evidence is presented that
the local abundance of star forming galaxies has been underestimated
and that the SFR only increases by a factor of 4 not 10 over this
range. We adopt this newer, more conservative value for our
calculations.  This factor of 4 is further reduced to $\approx 2.3$
when one moves into the $\Lambda$ dominated cosmology we subscribe to
in this paper (see Figure~\ref{fig13}).

\placefigure{fig13}

Taking this into account we can modify the local rate mentioned above
and arrive at an average rate at $z\sim2$ of $1.5 \times 10^{-4}$
SNe~II/year/Mpc$^{3}$. Combining this with the details of the light
curve calculations we ran simulations for both SNe~II-L and SNe~II-P
detections in the HDF. The results of these can be seen in
Figures~\ref{fig14} and \ref{fig15}. We have placed the graphs on the
same scale to illustrate how much easier it is to discover the
SNe~II-P in this type of search. The results of these simulations
predict that, on average, 0.47 SNe~II-L and 0.70 SNe~II-P will be
discovered in an HDF style search.

\placefigure{fig14}

\placefigure{fig15}

\subsection{Implications for SNe 1997ff and 1997fg}

Lacking either spectra, or significant photometric time-series information
for SNe 1997ff and 1997fg interpretation as Type Ia or II depends upon 
environmental and statistical factors which we summarize here.

Figures 16 and 17 provide histograms of the relative number of SNe detected
per magnitude interval in the HDF at host redshifts of z = 0.95 and 1.3 respectively
following from simulations as described in \S 5.1 and 5.2.
The statistical probability that any given event corresponds to a particular
SN type is a strong function of both magnitude at discovery and redshift.
For z = 0.95 (Fig.~16) it is interesting to note that for $m_I \; \sim$24 most
events are Type Ia -- this is broadly consistent with results of ground-based
searches which at this redshift and magnitude level are dominated by Type Ia
events.
At the fainter magnitudes to which our HDF search is sensitive the relative 
probability switches such that at $m_I$ = 26 both Types II-L and II-P are
statistically a factor of 2 more likely than a Type Ia.
For z = 1.3 (Fig.~17) similar behavior holds, although shifted to fainter
discovery levels in general and compressed in range due to our cutoff near $m_I \; \sim$27.5.

\placefigure{fig16}

\placefigure{fig17}

The general prevalence of expected Type II events relative to Type Ia at high z = 1.3
and faint magnitudes is counter-intuitive given results from ground-based searches
and the recognition that Type Ia events peak on average 2 magnitudes brighter
than Type IIs.
The assumed broad dispersion in peak magnitude for Type II SNe, coupled with their
higher intrinsic rate and greater relative UV flux {\em at maximum}
conspire to make the $HST$-HDF SNe search more sensitive to Type II SNe 
compared to Type Ia SNe {\em at the magnitude-redshift combination of SNe 1997ff and 1997fg.}

On the basis of environment for SN 1997ff where the host galaxy is an elliptical,
interpretation as a Type Ia is favored despite the overall statistical preference
as shown in Fig.~17 for Type II SNe at this magnitude redshift combination.

For SN 1997fg the host galaxy is an irregular which does not provide an environmental
restriction on SN type.
Fig.~16 would suggest that interpretation as a Type II event is favored at about
the 80\% level relative to the statistical probability of this being a Type Ia.

\section{Summary and Discussion}

We have shown that deep imaging {\em HST} data is well suited to the
detection and study of SNe at high redshift.  The long term stability
provided by {\em HST} allows for a straightforward quantification of
completeness limits that remain high to $m_I \; \gtrsim$ 27 for high
redshift galaxy hosts in these data, and for which the dependence on
host galaxy redshift is well quantified.

A search of two epochs of HDF data in the F814W filter is sensitive to
SNe~Ia and II to z $\sim$ 1.8.  We detected two supernovae, one of which is
most probably a SNe~Ia based
primarily on its late type host galaxy;
the other with a (spectroscopic)
redshift of 0.952 for the host galaxy and $m_I \; \sim$ 26.0 may be
either a SN~Ia or SN~II event, but factoring in the early host galaxy type
and relative detection rates developed in \S 5 statistically favors a
Type II interpretation.

Our two SNe detections from a two epoch search of the HDF are
consistent with general expectations for the cosmic rates of Type Ia
and II events.  These small number statistics do not yet support
placing meaningful constraints on different cosmologies, or
assumptions regarding Type Ia formation for which the parameter space
of possible rates at modest z $\sim$ 1 - 2 is still relatively open.
We have demonstrated with current technology a sensitivity to
detecting SNe to z $\sim$ 1.8 and note that with the much improved
``discovery space" (product of area and efficiency) of the Advanced
Camera for Surveys (to be installed on HST in 2000) in the 0.8 - 1.0
$\mu$ range relative to WFPC2 that programs to return a statistically
robust sample of both Type Ia and II events to z $\sim$ 2 will be
feasible.

\acknowledgments

We thank Andy Fruchter for numerous discussions related to data
reductions of the HDF data; illuminating exchanges are also
appreciated with Rodrigo Ibata. Discussions with Greg Aldering
concerning the SFR were very enlightening as were conversations with
Doug Leonard on the spectra of SNe~II.
Massimo Turatto kindly provided digital copies of their SN II light curve templates.
Alberto Fern\'{a}ndez-Soto and
Ken Lanzetta kindly provided access to their data base on HDF galaxy
redshifts and properties well in advance of publication and provided
associated discussion.  We thank Mark Dickinson for general discussion
of the HDF and photometric redshift determinations. We thank Bob
Williams for encouragement and for initiating the HDF observations.
R.L.G. acknowledges the hospitality of the Astronomy and Astrophysics
Department at UC Santa Cruz where much of this work was done, the
ST ScI for a sabbatical leave, and support via GO-6473.01-95A. Calculations presented in this paper
were performed at the National Energy Research Supercomputer Center;
P.E.N. acknowledges support under DoE 76SF0098 and by the Director,
Office of Computational and Technology Research of the U.S. DoE.

\pagebreak 


\clearpage

\figcaption[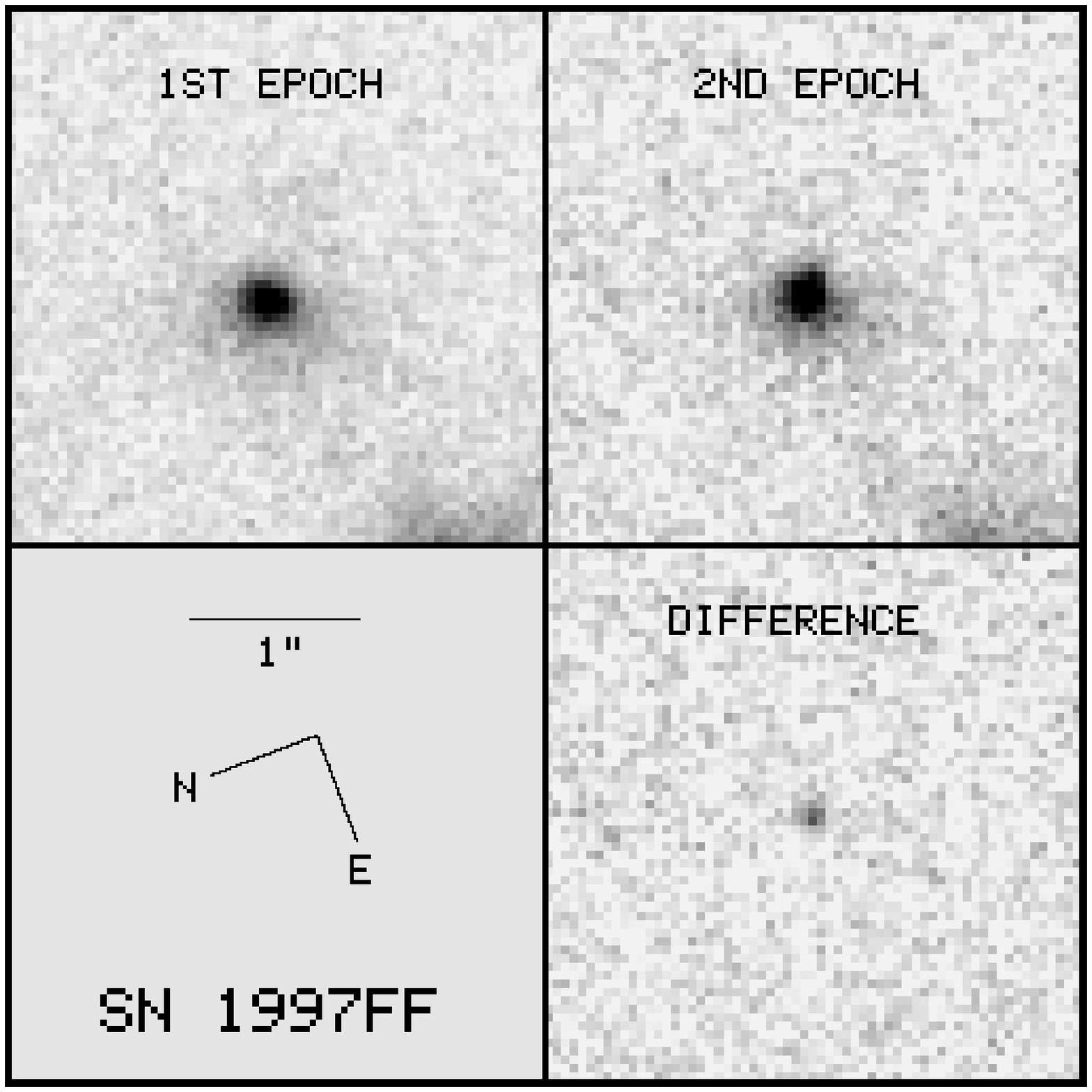]{First and second epoch direct images of SN 1997ff in galaxy
4-403.0 are shown in the upper panels.  The lower right panel shows
the difference image (second minus first epoch) in which a point
source residual at a magnitude of $m_I \sim $ 27.0 corresponding to
the supernova can be seen.\label{fig1}}

\figcaption[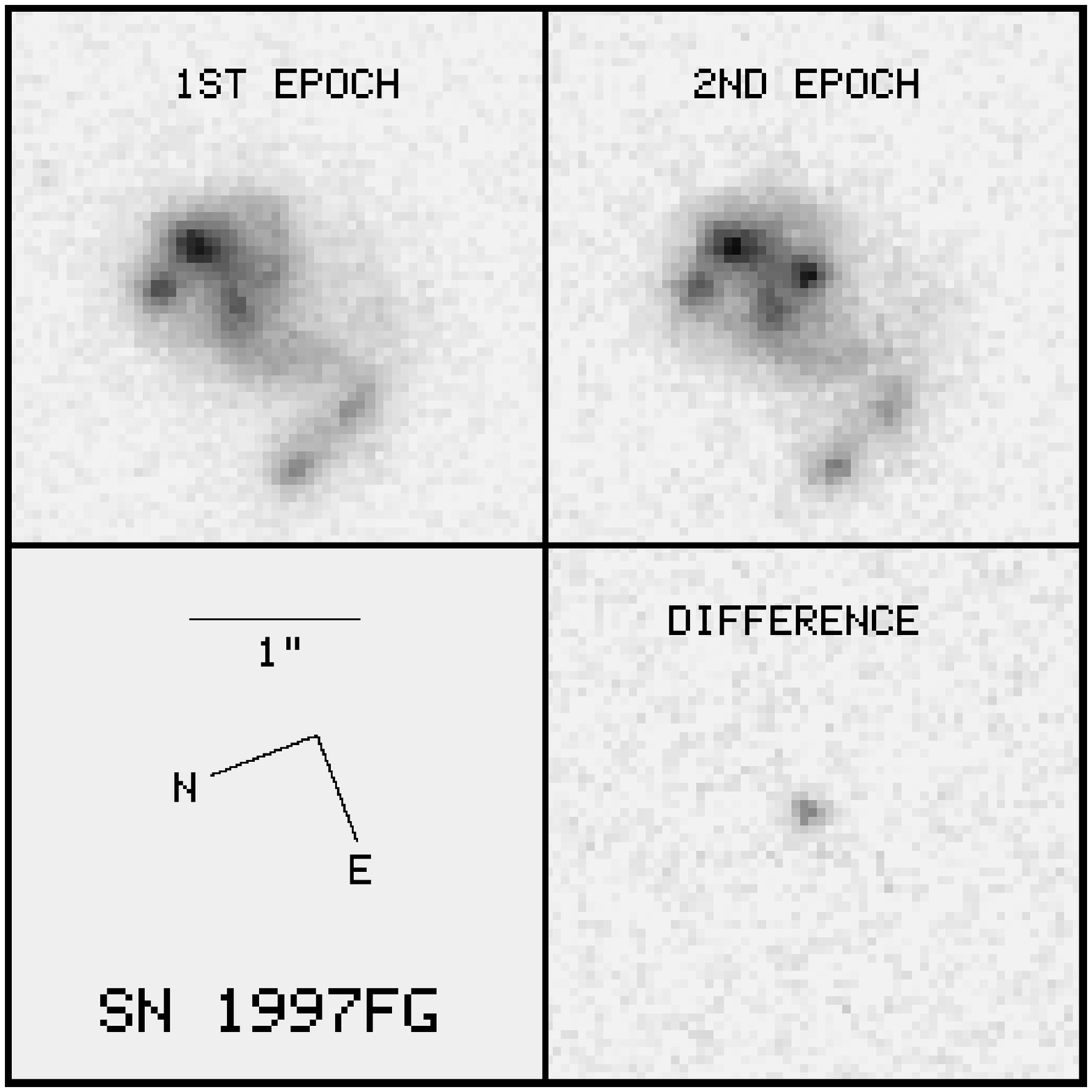]{The first and second epoch direct images of SN 1997fg in
galaxy 3-221.0, and the difference image showing the $m_I \sim$ 26.0
point source corresponding to the supernova.\label{fig2}}

\figcaption[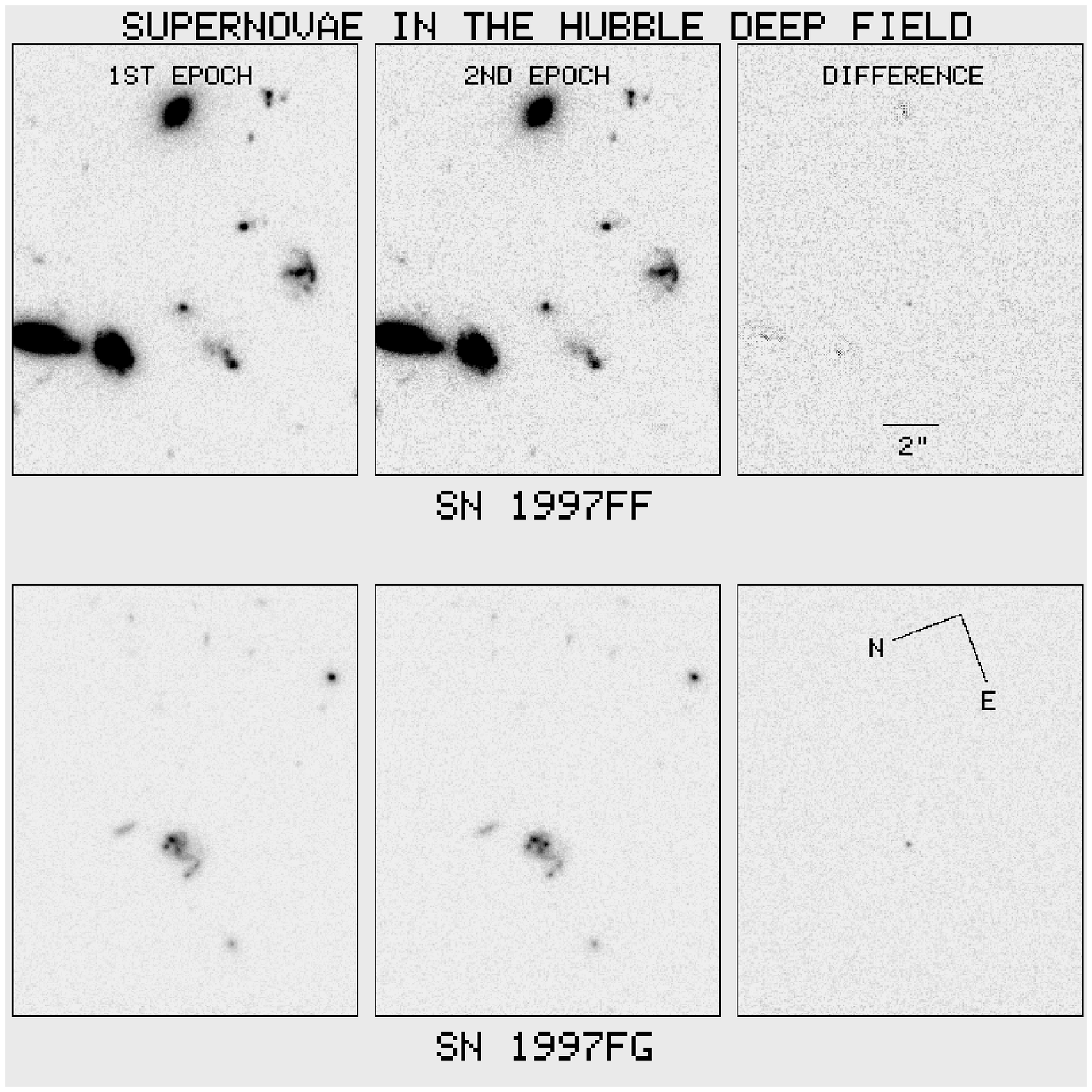]{Direct and difference images over a wider area than in Figs 1
\& 2 for both SN 1997ff and SN 1997fg.\label{fig3}}

\figcaption[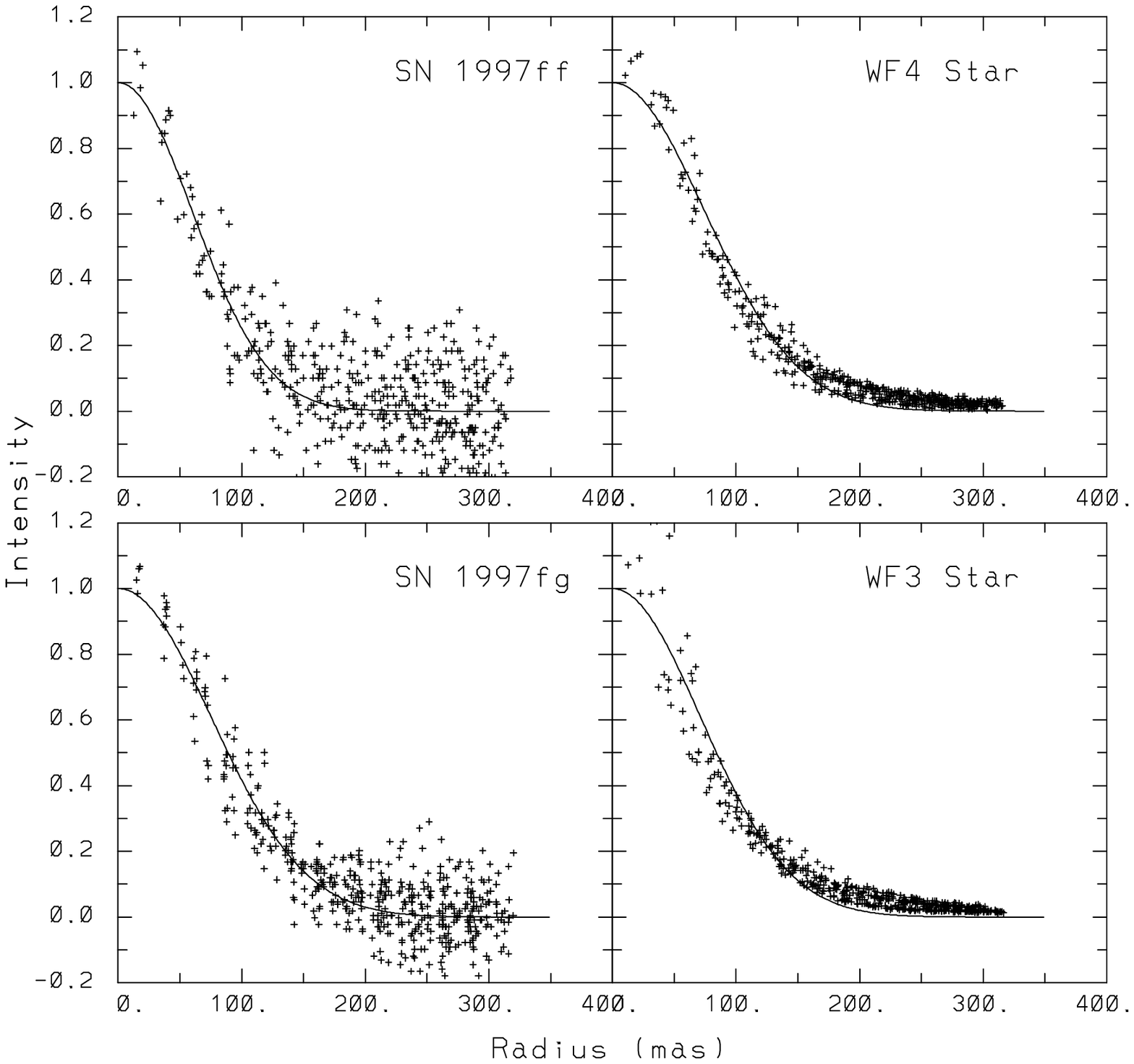]{Radial plots and associated Gaussian profile fits for (left
panels) SN 1997ff and 1997fg, and (right panels) stars from the same
CCD as the respective supernovae.\label{fig4}}

\figcaption[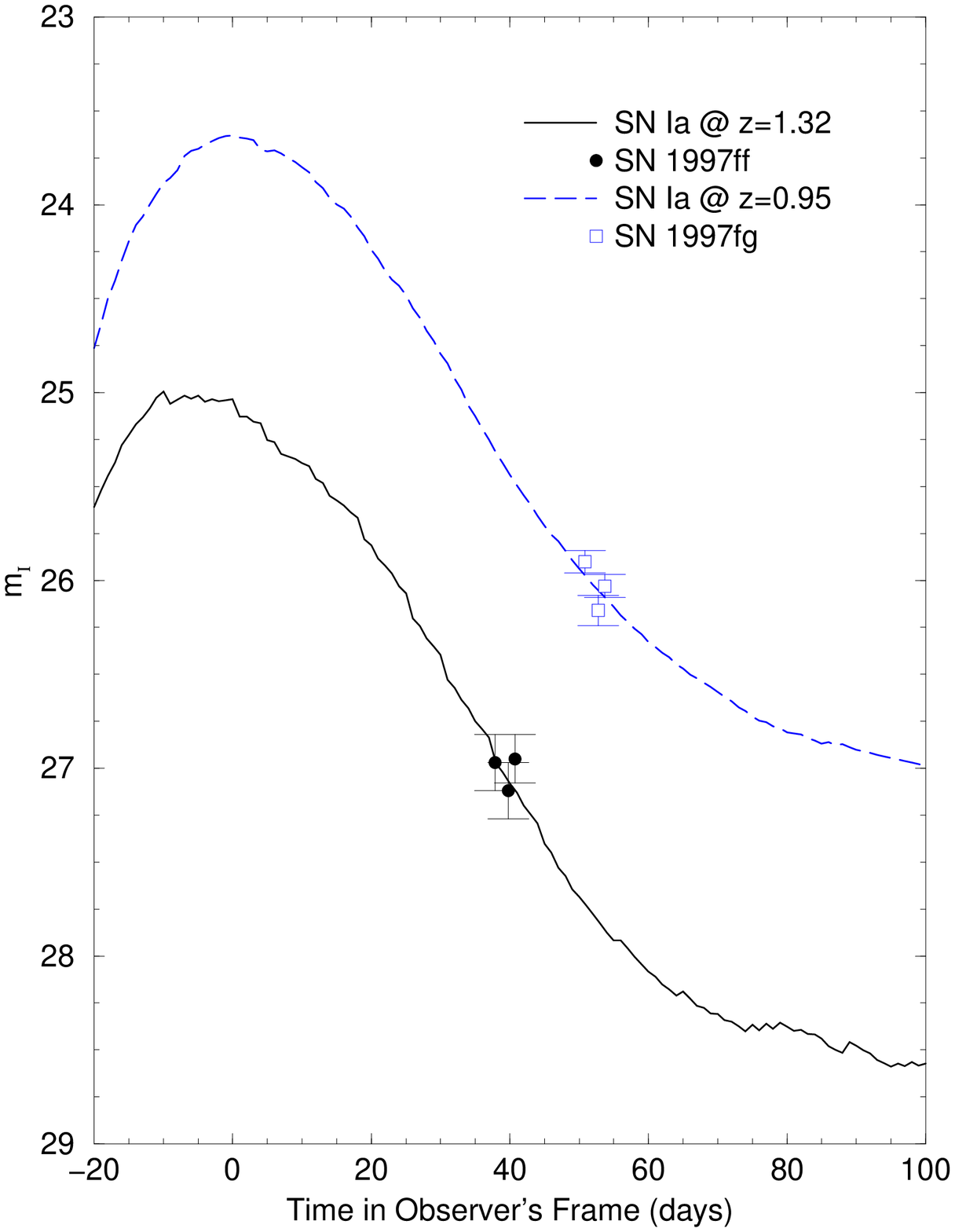]{The daily magnitudes derived for each of SN 1997ff and 1997fg
are overplotted on a Type Ia supernova light curve that has been
properly shifted in magnitude and time dilated for the assumed
redshift of the host galaxy. Time is with respect to maximum light in
the $B$-band.\label{fig5}}

\figcaption[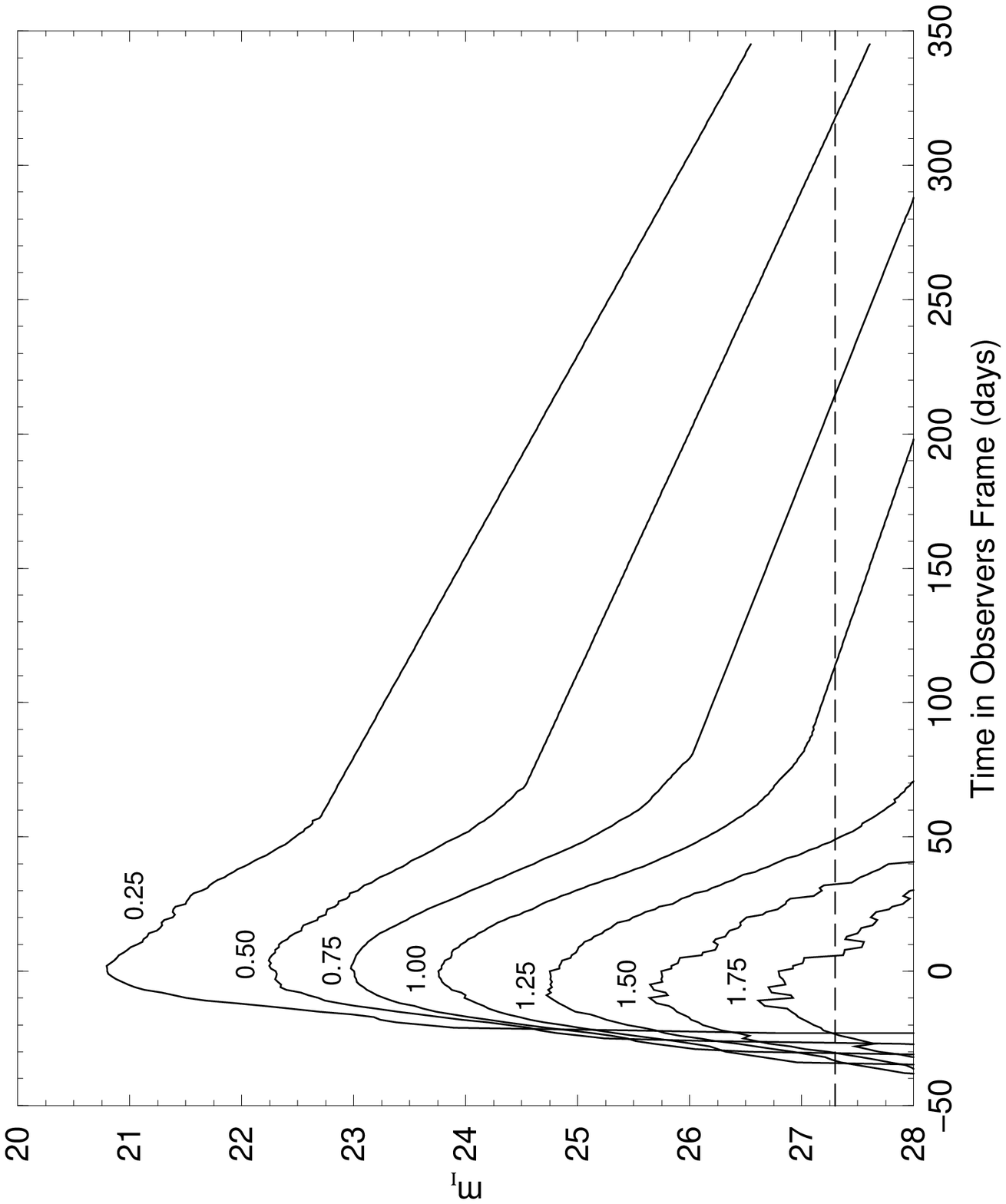]{Light curves for Type Ia SNe derived for z = 0.25 (top) to z
= 1.75 (bottom) in I-band apparent magnitudes for an event of average
intrinsic brightness.  A fiducial dashed line at $m_I$ = 27.3 shows
the $\sim$50\% completeness limit for z = 1.2 galaxies in our HDF
data.
\label{fig6}}

\figcaption[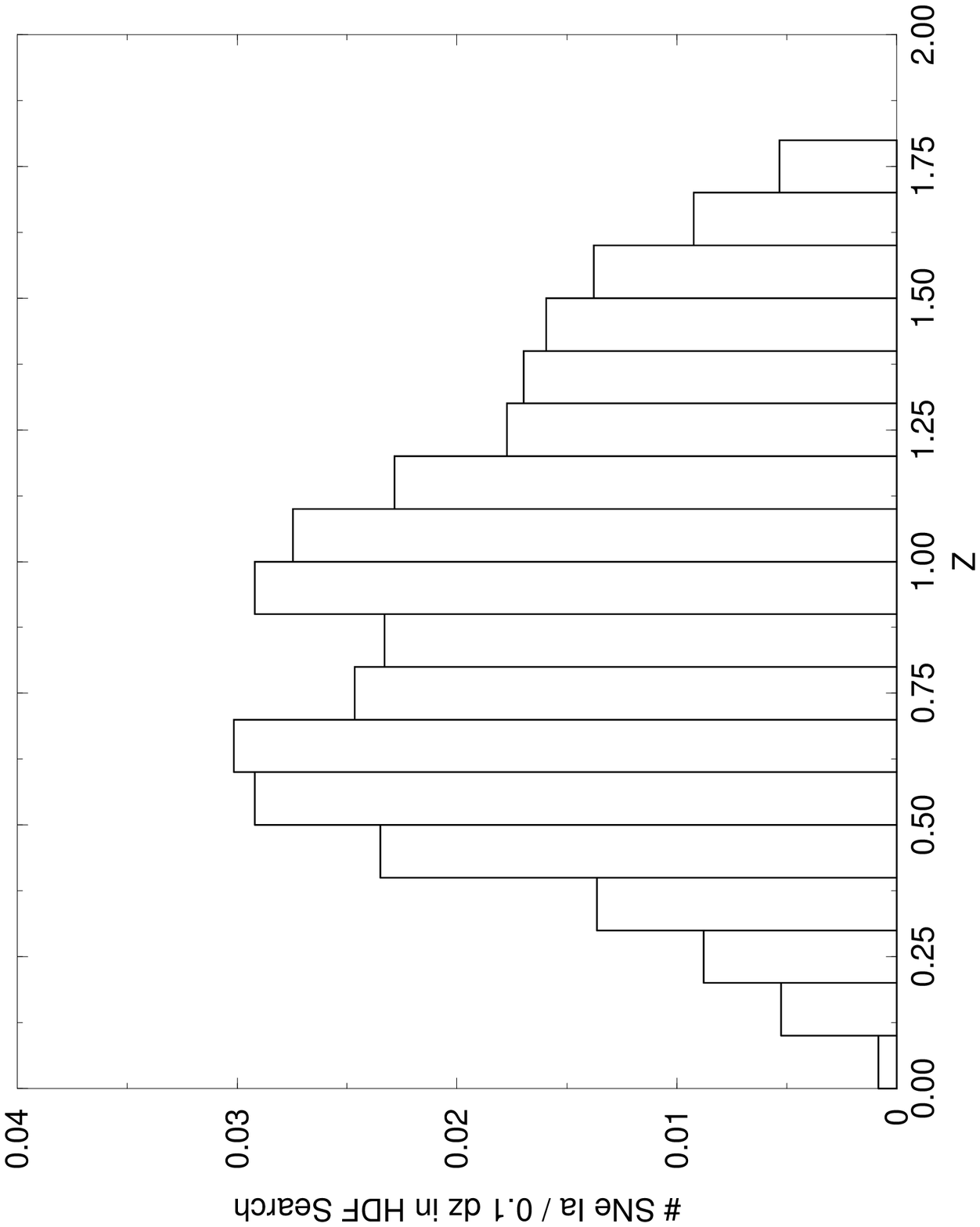]{Expected number of SNe Ia events in the two epochs of the HDF
data per unit redshift based on assuming a constant rate per unit
volume with redshift (at the Pain {\em et al.} 1997 z $\sim$ 0.4
value) and the assumed $\Omega_M$ = 0.28, $\Omega_{\Lambda}$ = 0.72,
$h = 0.633$ cosmology.
\label{fig7}}

\figcaption[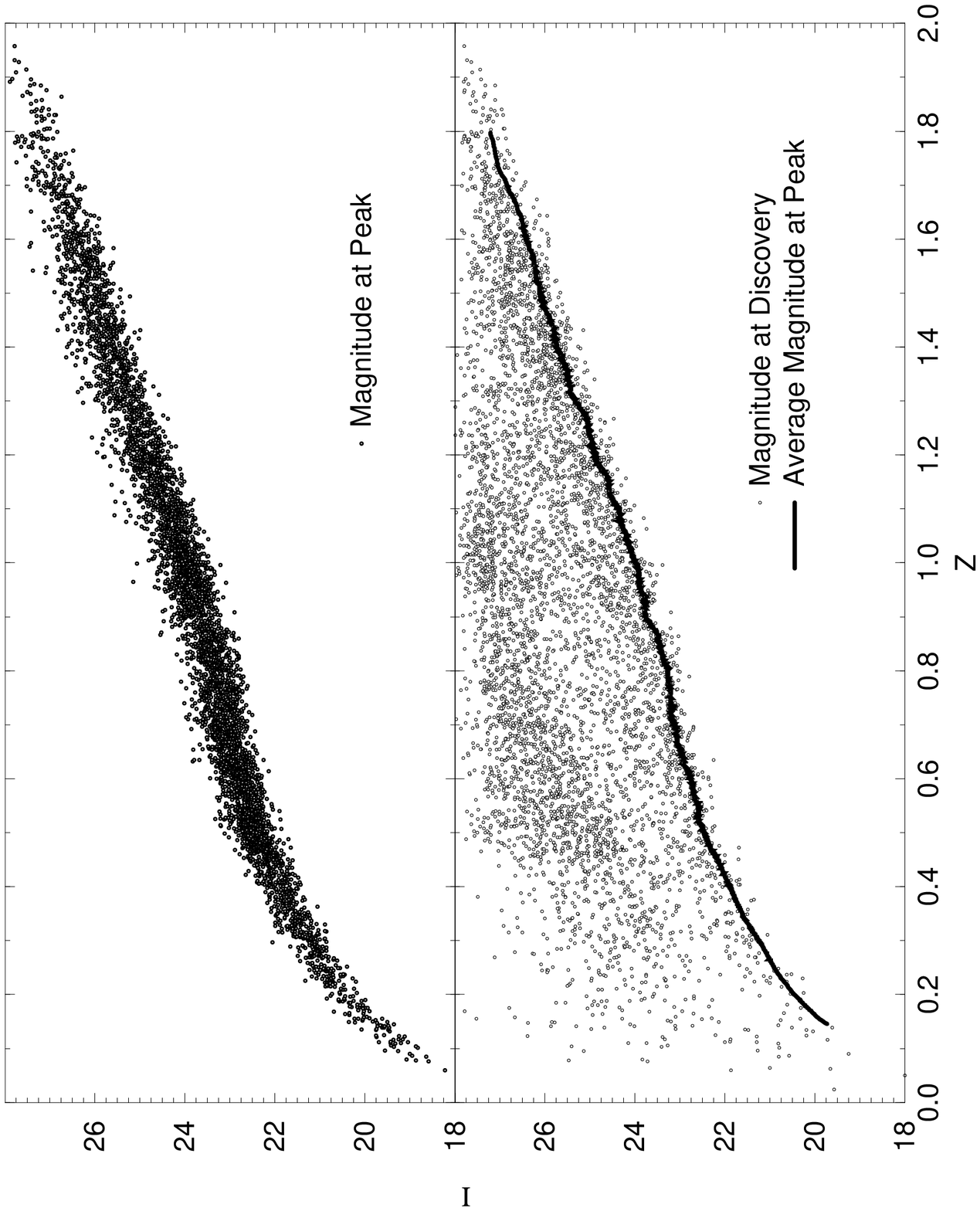]{(Upper) Apparent $m_I$ magnitudes at peak plotted against
redshift for the 5000 SNe~Ia generated in our Monte Carlo simulation
of an HDF search for an $\Omega_M$ = 0.28, $\Omega_{\Lambda}$ = 0.72,
$h = 0.633$ cosmology.  (Lower) Apparent $m_I$ magnitudes at discovery
plotted against redshift for the same simulation.  The solid line
shows for comparison the ridge line of input peak magnitudes.
\label{fig8}}

\figcaption[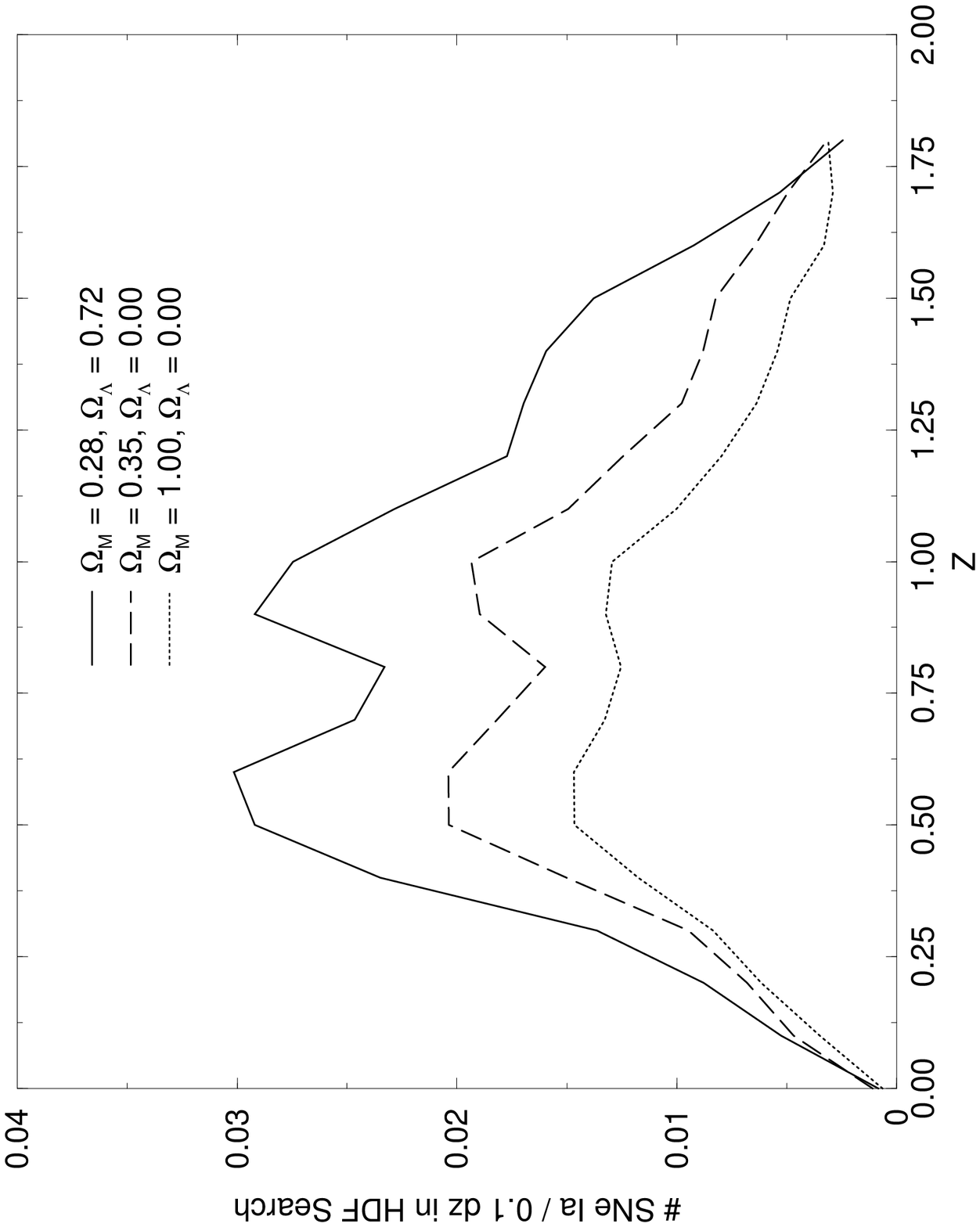]{The expected number of SNe Ia events in the two epochs of the
HDF data per unit redshift are contrasted for three different
cosmologies: {\em Solid line}: a cosmological constant dominated
universe, {\em Dashed line}: an open, matter only universe, and {\em
Dotted line}: a matter only, critical density universe.
\label{fig9}}

\figcaption[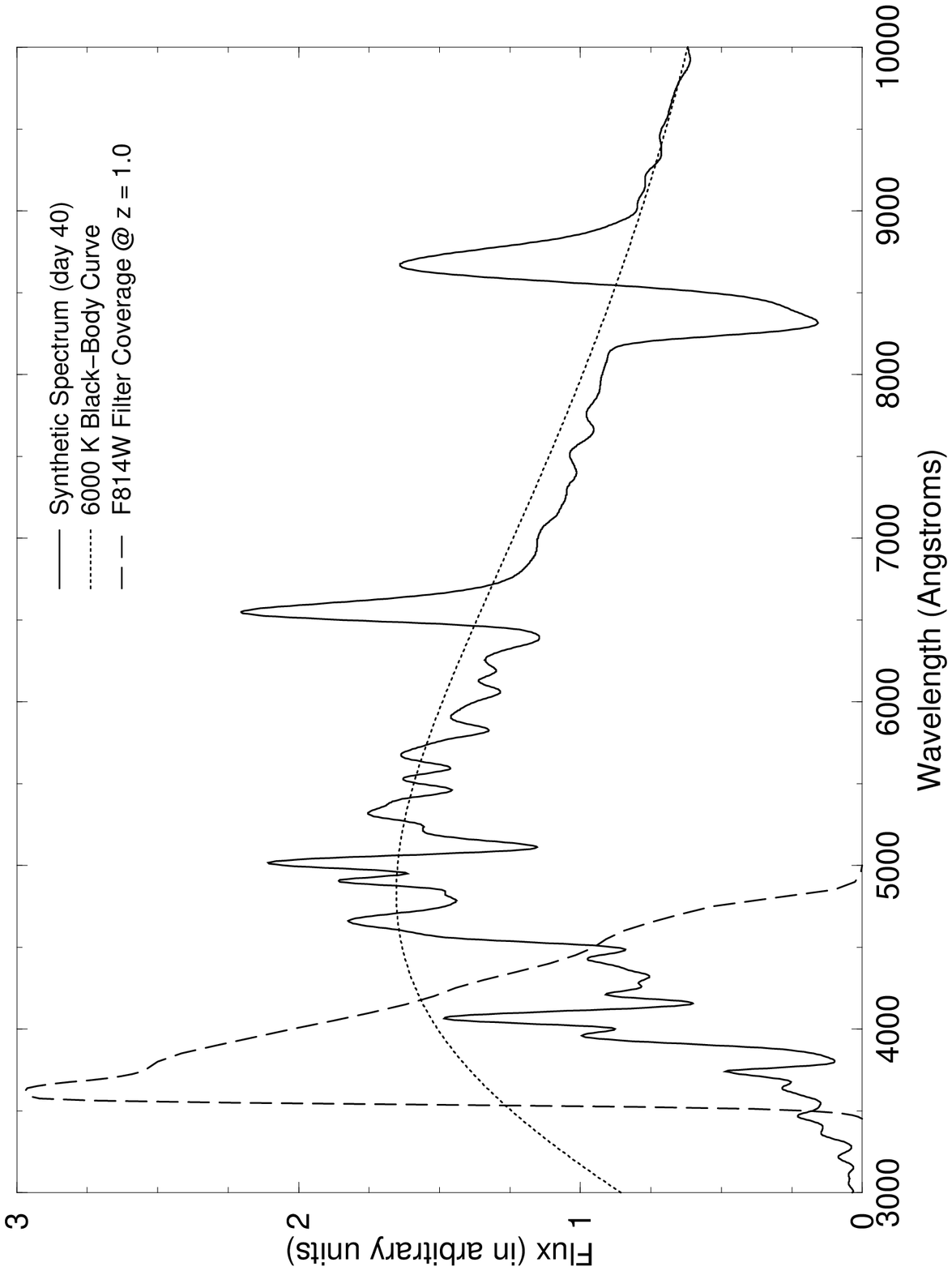]{Comparison of synthetic SNe II spectrum (solid curve) with a
representative blackbody (dotted curve) matching the optical.  The
dashed curve shows the wavelength origin that F814W passes for a
modest z = 1.0 redshift -- for typical cases the UV flux is a small
fraction of that predicted from common blackbody representations.
\label{fig10}}

\figcaption[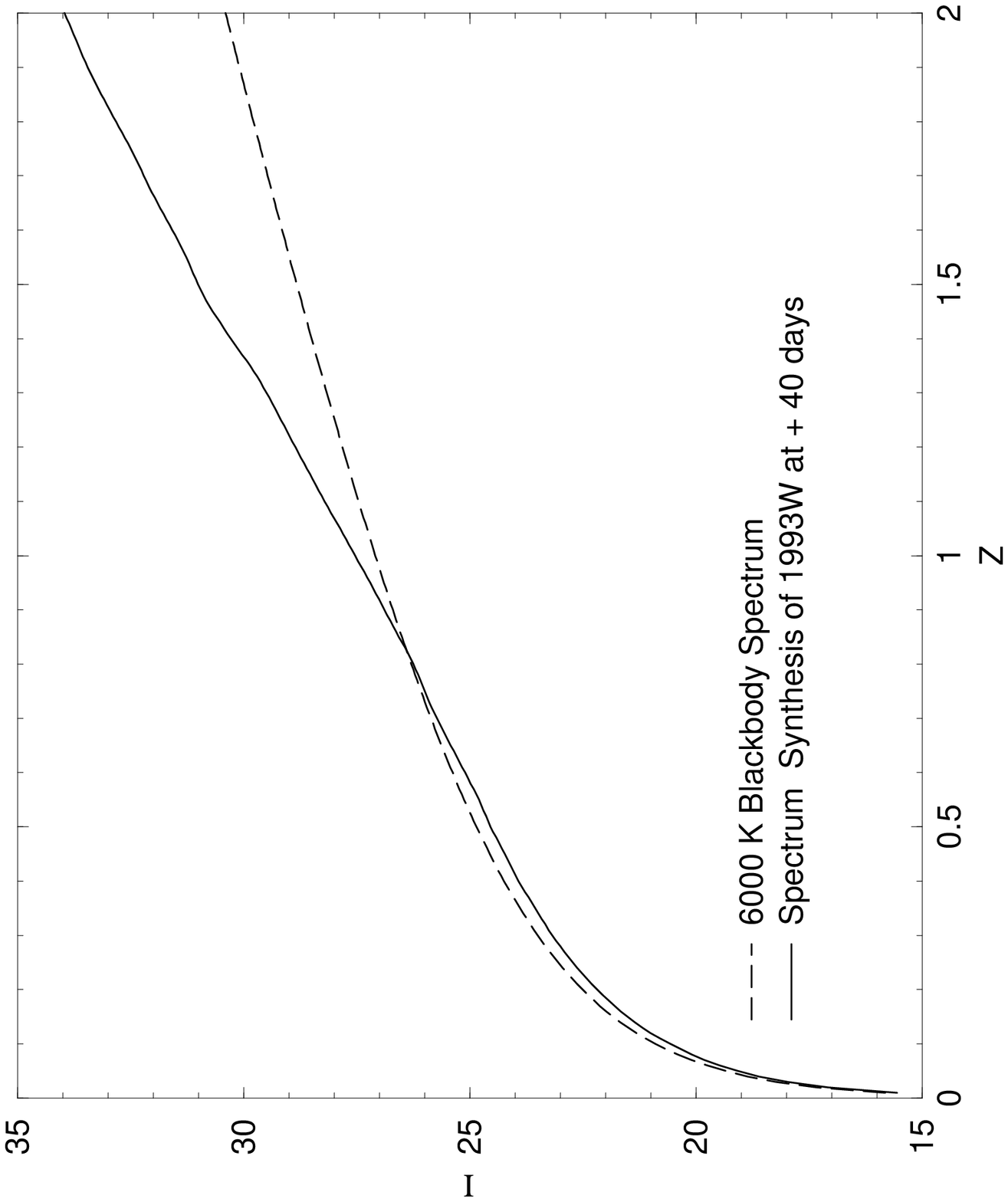]{I-band magnitude comparison of representative blackbody
(dashed line) and a spectrum synthesis model for a SNe II-P
mid-plateau phase 40 days after outburst.  At redshifts beyond z
$\sim$ 0.8 predicted magnitudes for Type II-P events drop rapidly in
the spectrum synthesis model due to suppressed UV flux.
\label{fig11}}

\figcaption[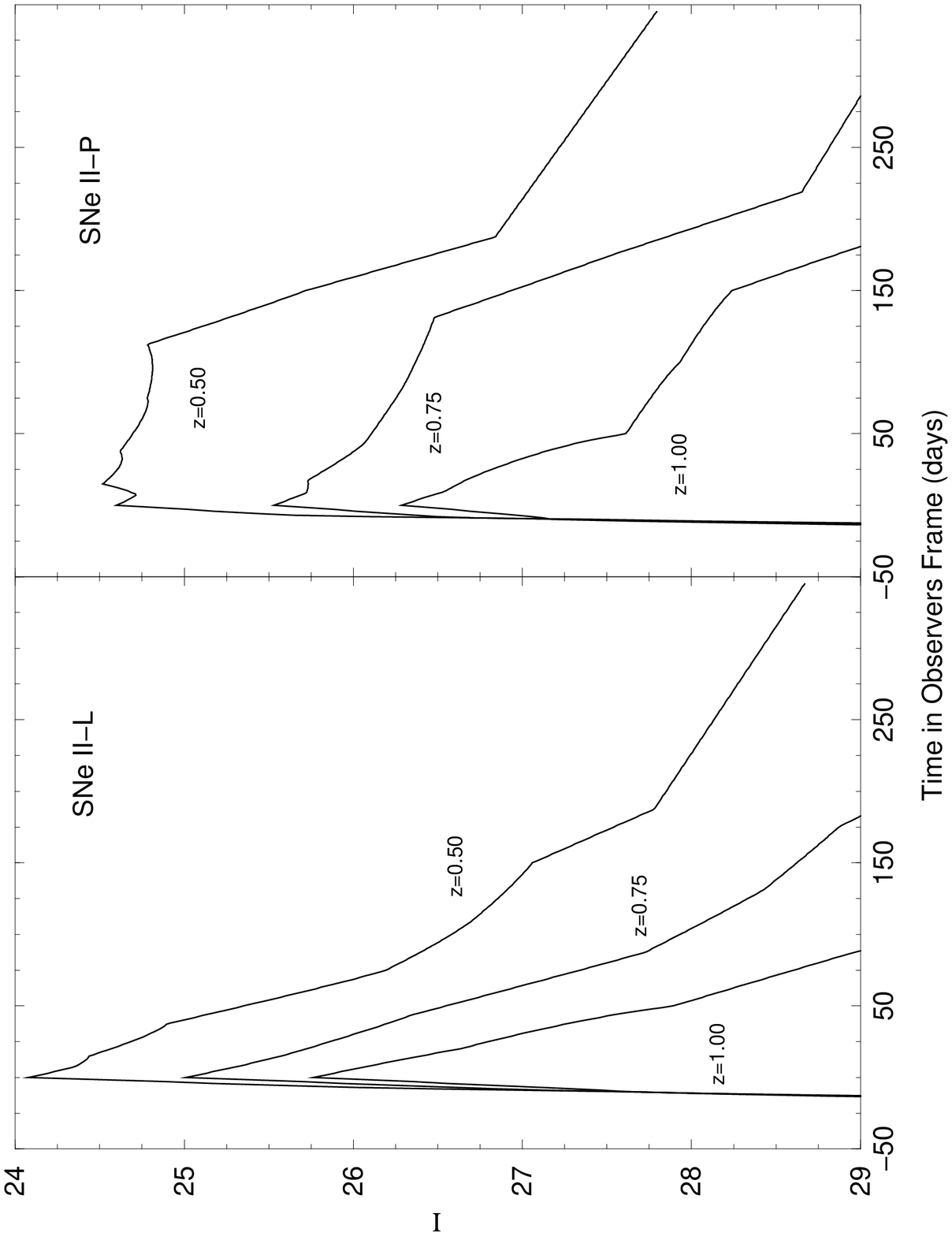]{Representative light curves for Type II SNe at redshifts of
0.5, 0.75 and z = 1.0 (compare with Ia light curves in Fig. 6).
\label{fig12}}

\figcaption[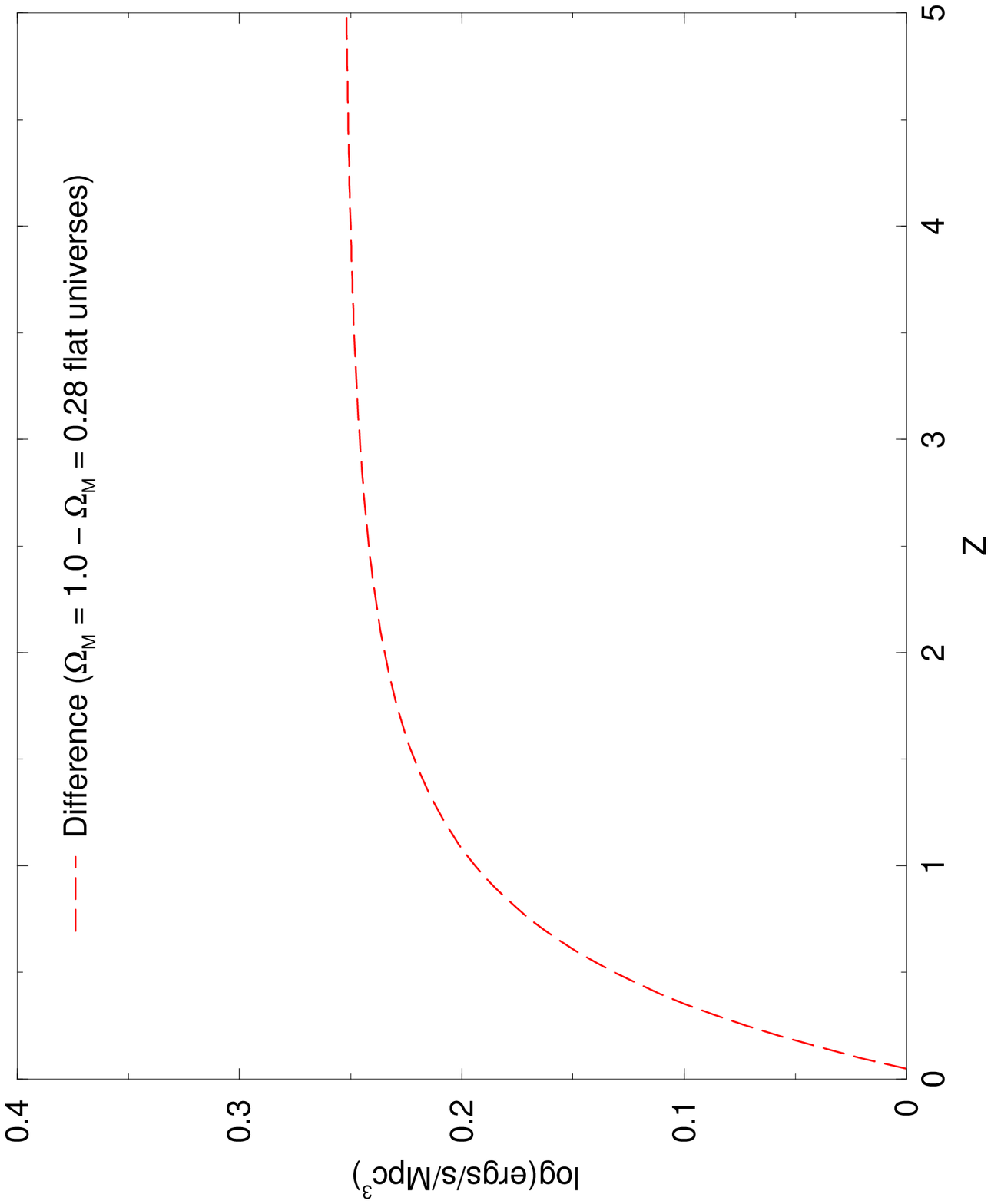]{The difference in luminosity density arising from the assumed
Madau, Pozzetti, \& Dickinson (1998) $\Omega_M$ = 1.0 and our assumed
standard flat, but $\Lambda$ dominated universe.\label{fig13}}

\figcaption[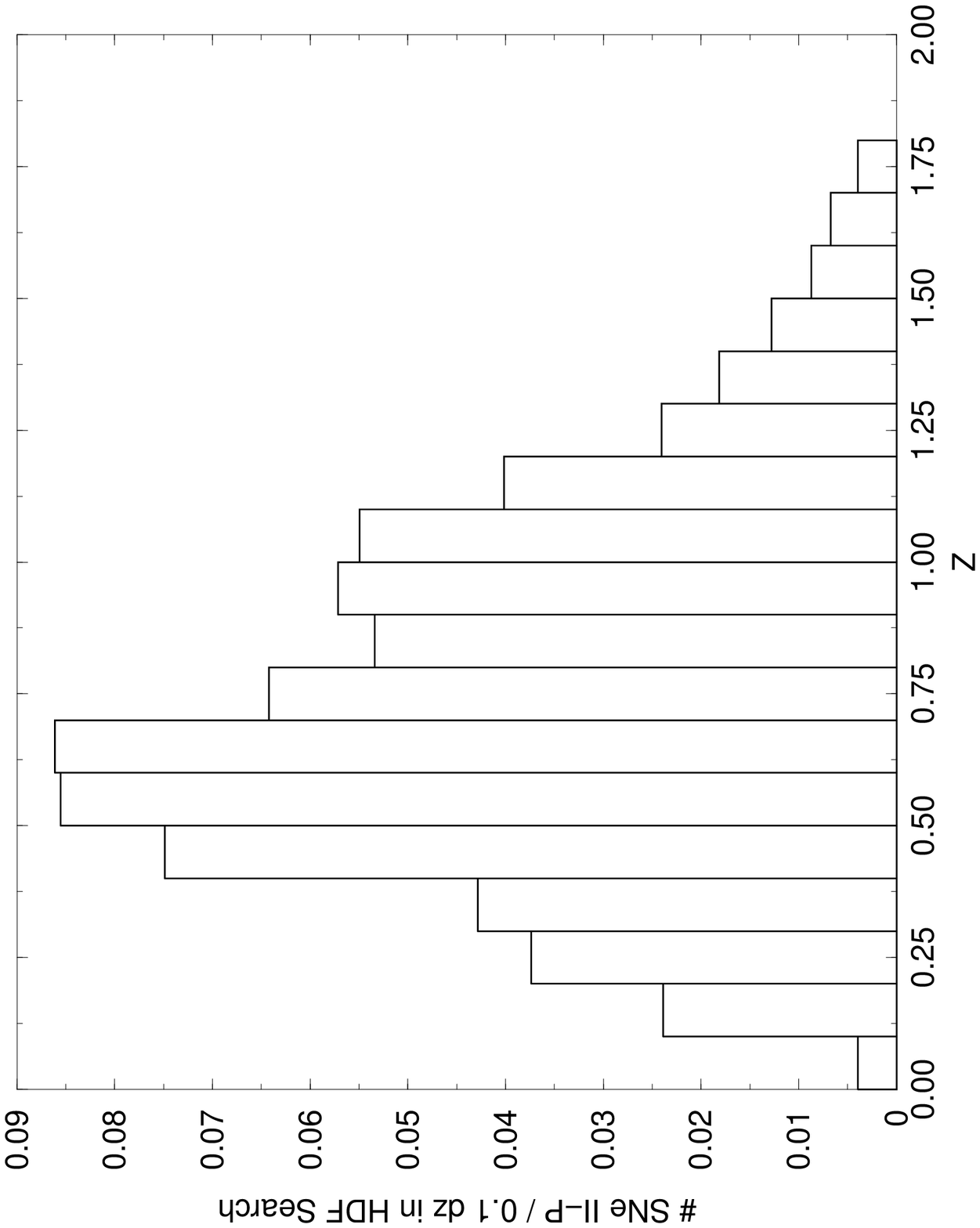]{Expected number of SNe II-P events in the two epochs of HDF
data per unit redshift based on assuming a constant rate (at 2.3
$\times$ local -- see text) per unit volume with redshift.
\label{fig14}}

\figcaption[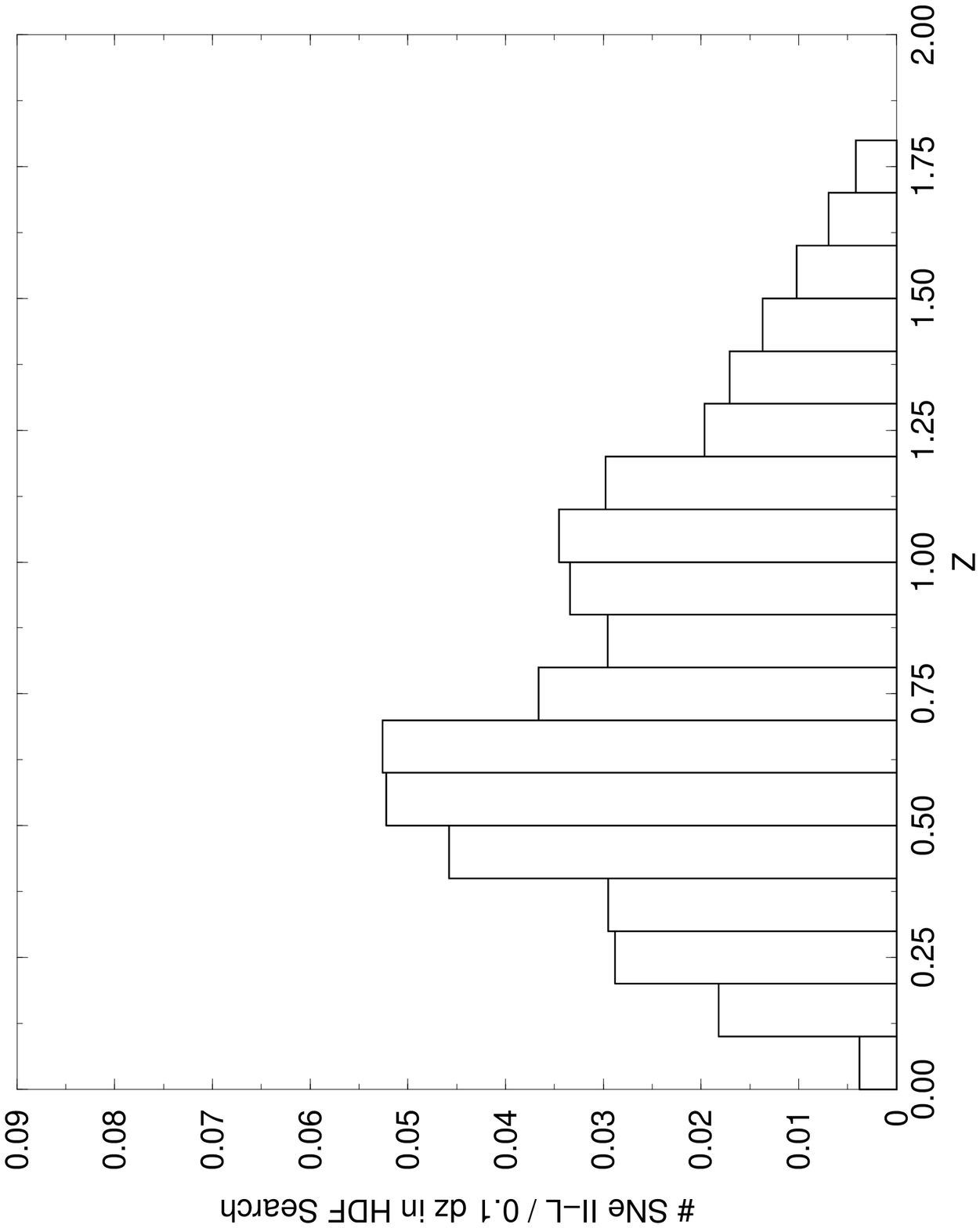]{Same as for Fig. 12, but for SNe II-L.
\label{fig15}}

\figcaption[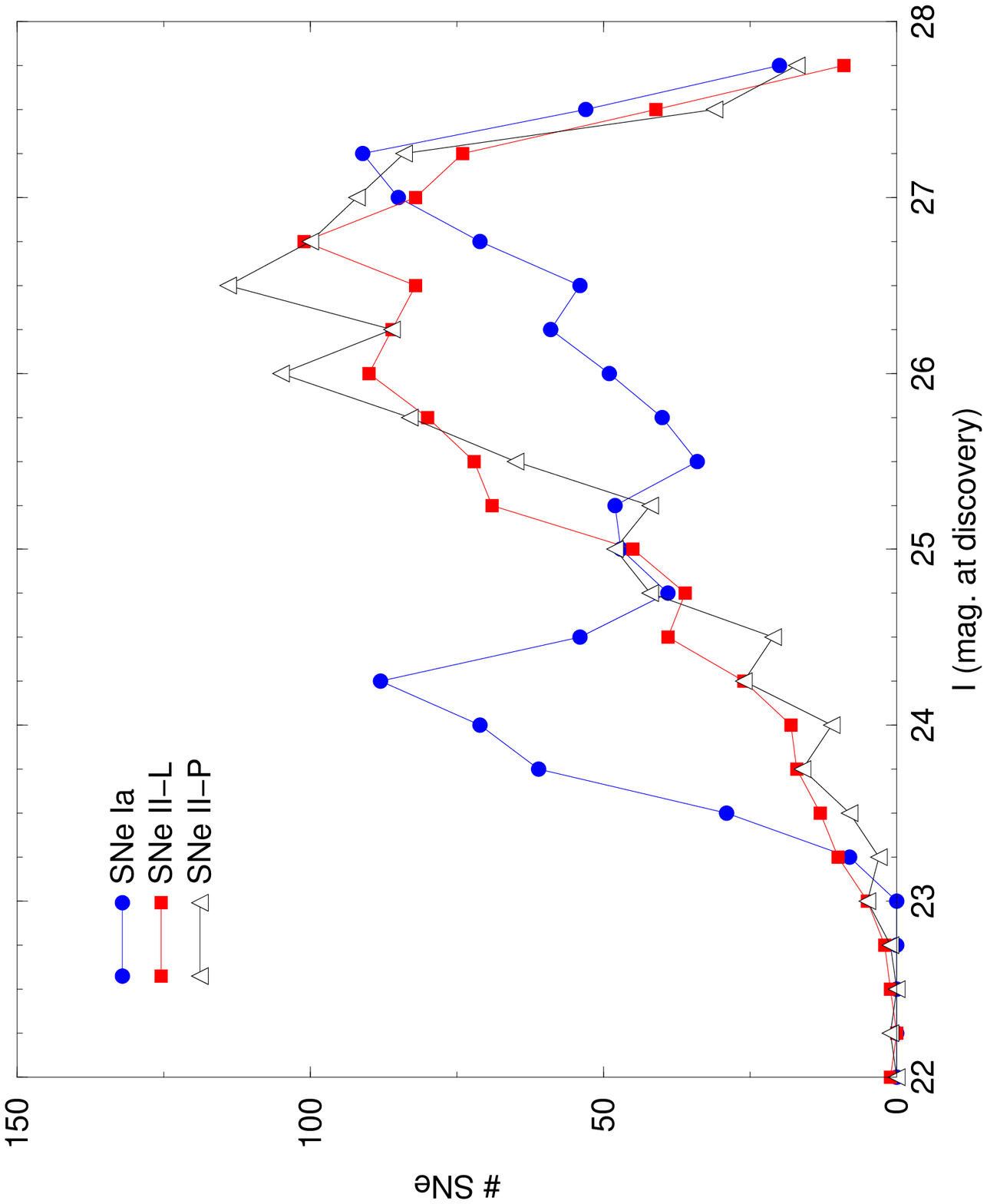]{A histogram of the discovery magnitudes given our simulations
for the three types of SNe at a $z=0.95$.
\label{fig16}}

\figcaption[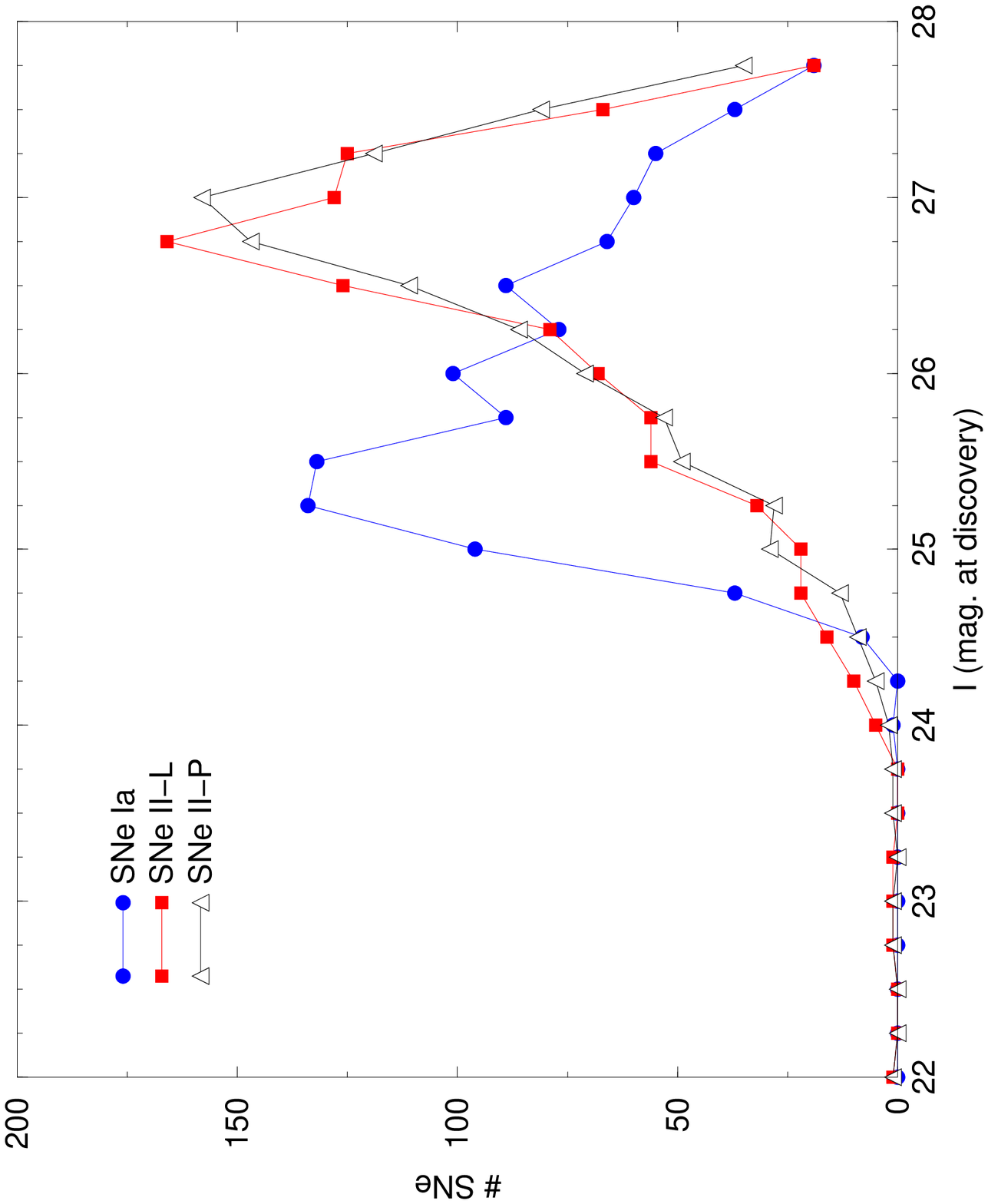]{Same as for Fig.~16, but for a $z=1.3$.
\label{fig17}}

\clearpage

\begin{deluxetable}{ccccrcc}
\tablecaption{First and Second Epoch Data Summary}
\tablehead{\colhead{ } & \colhead{Mean epoch} &
\colhead{Number} & \colhead{Number} & \colhead{Total Exp.} 
& \colhead{ } &\colhead{ } \nl
\colhead{Filter} & \colhead{(MJD)} &
\colhead{Frames} &
\colhead{Dithers} &
\colhead{Time (s)} &
\colhead{$T_{exp}$/frame (s)} &
\colhead{Sky (e-/s)}
}
\startdata
F814W & 50075.5594 & 58 & 10 & 123600 & 1100-3100 & 0.0445 \nl
F814W & 50807.4587 & 36 & 17 &  63000 & 1600,1900 & 0.0448 \nl
F300W & 50807.4819 & 18 & 17 &  27300 & 1100,1600 & 0.0038 \nl

\enddata
\label{tab1}
\end{deluxetable}

\clearpage

\begin{deluxetable}{ccrrr}
\tablecaption{Comparison of PSF FWHM and Background Noise with V. 2
Drizzle Release}  
\tablehead{
\colhead{Filter} &
\colhead{Chip} &
\colhead{\% diff. noise\tablenotemark{a}} &
\colhead{\% diff. FWHM\tablenotemark{b}} &
\colhead{Mag. gain\tablenotemark{c}} 
}

\startdata
F300W & PC1 & -21.8 & -7.9 & 0.28 \nl
F300W & WF2 & -6.9 & +0.7 & 0.07 \nl
F300W & WF3 & -9.3 & -6.8 & 0.16 \nl
F300W & WF4 & -1.1 & -7.1 & 0.09 \nl
F814W & PC1 & -6.0 & -6.9 & 0.13 \nl
F814W & WF2 & -5.4 & -5.1 & 0.11 \nl
F814W & WF3 & +0.4 &  0.0 & -0.00 \nl
F814W & WF4 & -1.4 & -9.6 & 0.10 \nl
\enddata

\tablenotetext{a}{Noise levels were evaluated as the standard
deviation of pixel values in the same regions of both reductions
dominated by background and corrected for the effects of over-sampling
(correlated noise), negative values reflect lower noise levels in the
current reductions.}  \tablenotetext{b}{PSF widths were measured for
an isolated bright star (or closest approximation in case of PC data)
using the IRAF-imexamine task and a radial fit to report the enclosed
energy statistic for width; negative values reflect sharper PSFs in
these reductions.}  \tablenotetext{c}{Expected point-source limiting
magnitude gain of new reductions (for original epoch only) as 2.5 $\log _{10}$(1 - fractional
noise - fractional FWHM difference).}
\label{tab2}
\end{deluxetable}

\clearpage

\begin{deluxetable}{cccccrl}
\tablecaption{Top False Alarms in Control Experiments}
\tablehead{
\colhead{Chip} &
\colhead{Epoch} &
\colhead{Type\tablenotemark{a}} &
\colhead{$m_I$} &
\colhead{mag-err} &
\colhead{sharp\tablenotemark{b}} &
\colhead{host} 
}
\startdata
WF4 & 1 & sim & 27.853 & 0.367 & -0.072 & $m_I$ = 20.9 galaxy \nl
WF3 & 1 & sim & 27.860 & 0.235 & -0.036 & none \nl
WF3 & 2 & e-o & 27.864 & 0.201 & -0.187 & $m_I$ = 21.4 galaxy \nl
WF4 & 1 & e-o & 27.895 & 0.230 & -0.043 & none \nl
WF2 & 2 & e-o & 27.898 & 0.328 & -0.168 & $m_I$ = 21.4 galaxy \nl

\enddata

\tablenotetext{a}{Type of control experiment, ``e-o'' stands for the
even minus odd numbered frame epochs, and ``sim'' for the simulation of
noise.}  \tablenotetext{b}{The sharpness statistic reported by
DAOPHOT, stellar sources should have values near zero.}
\label{tab3}
\end{deluxetable}

\clearpage

\begin{deluxetable}{ccccrl}
\tablecaption{Top Potential SN Candidates}
\tablehead{
\colhead{Chip} &
\colhead{Epoch} &
\colhead{$m_I$} &
\colhead{mag-err} &
\colhead{sharp} &
\colhead{host, ID}
}
\startdata
WF3 & 2 & 26.014 & 0.093 & 0.115 & z=0.952, SN 1997fg \nl
WF4 & 2 & 26.957 & 0.079 & -0.018 & z=1.32, SN 1997ff \nl
WF2 & 2 & 27.100 & 0.171 & 0.108 & $m_I$ = 18.7 star \nl
WF3 & 1 & 27.351 & 0.284 & 0.027 & z=0.52, 3-404.2 \nl
WF2 & 1 & 27.453 & 0.262 & -0.043 & $m_I$ = 18.7 star \nl
WF3 & 2 & 27.512 & 0.191 & -0.153 & $m_I$ = 18.9 star \nl
WF3 & 1 & 27.642 & 0.297 & 0.114 & $m_I$ = 18.9 star \nl

\enddata

\label{tab4}
\end{deluxetable}

\clearpage

\begin{deluxetable}{ccccc}
\tablecaption{Completeness Limits (\%)}
\tablehead{
\colhead{Magnitude} &
\colhead{z =0.2\tablenotemark{a}} &
\colhead{z = 0.7} &
\colhead{z = 1.2} &
\colhead{z = 1.7}
}
\startdata
23.7 & 72 & 90 & 98 & 100 \nl
24.9 & 53 & 83 & 98 & 100 \nl
25.2 & 47 & 80 & 97 & 100 \nl
25.5 & 41 & 76 & 96 & 100 \nl
25.8 & 28 & 72 & 92 & 100 \nl
26.1 & 22 & 62 & 89 & 100 \nl
26.4 & 13 & 56 & 84 & 100 \nl
26.7 & 9 & 47 & 78 & 99 \nl
27.0 & 8 & 38 & 72 & 94 \nl
27.3 & 5 & 27 & 56 & 77 \nl
27.6 & 2 & 10 & 26 & 42 \nl
27.9 & 0 & 3 & 8 & 7 \nl

\enddata

\tablenotetext{a}{Galaxy redshift range, to which artificial stars
were added as described in text, of listed central value $\pm$0.2.}
\label{tab5}
\end{deluxetable}

\clearpage

\begin{deluxetable}{ccc}
\tablecaption{Time Resolved Magnitudes}
\tablehead{
\colhead{Date (UT) Dec. 1997} &
\colhead{SN 1997ff} &
\colhead{SN 1997fg} 
}
\startdata
23.83 & 26.97 $\pm$ 0.15 & 25.90 $\pm$ 0.06 \nl
25.78 & 27.12 $\pm$ 0.15 & 26.16 $\pm$ 0.08 \nl
26.74 & 26.95 $\pm$ 0.13 & 26.03 $\pm$ 0.06 \nl
Mean\tablenotemark{a} & 27.00 $\pm$ 0.08 & 26.02 $\pm$ 0.04 \nl

\enddata

\tablenotetext{a}{The mean magnitude follows from PSF fit analysis
over the full data set.}
\label{tab6}
\end{deluxetable}

\clearpage

\end{document}